\documentclass[aip,jcp,amsmath,amssymb,floatfix,reprint,citeautoscript]{revtex4-2}

\usepackage[utf8]{inputenc}
\usepackage{graphicx}
\usepackage{booktabs}
\usepackage{wrapfig}
\usepackage{bibunits}
\usepackage[colorlinks,allcolors=black,citecolor=blue,urlcolor=blue]{hyperref}
\usepackage{ae}
\usepackage{amsmath}
\usepackage{physics}

\DeclareMathOperator{\arctan2}{arctan2}

\usepackage{silence}
\graphicspath{{figures/}}

\defaultbibliography{paper,revtex-control}
\defaultbibliographystyle{aipnum4-2}

\begin{document}

\def\mytitle{Electronic structure of the solvated benzene radical anion}
\title{\mytitle}

\author{Krystof Brezina}
\affiliation{
Charles University, Faculty of Mathematics and Physics, Ke Karlovu 3, 121 16 Prague 2, Czech Republic
}
\affiliation{
Institute of Organic Chemistry and Biochemistry of the Czech Academy of Sciences, Flemingovo nám. 2, 166 10 Prague 6, Czech Republic
}

\author{Vojtech Kostal}
\affiliation{
Institute of Organic Chemistry and Biochemistry of the Czech Academy of Sciences, Flemingovo nám. 2, 166 10 Prague 6, Czech Republic
}

\author{Pavel Jungwirth}
\affiliation{
Institute of Organic Chemistry and Biochemistry of the Czech Academy of Sciences, Flemingovo nám. 2, 166 10 Prague 6, Czech Republic
}

\author{Ondrej Marsalek}
\email{ondrej.marsalek@mff.cuni.cz}
\affiliation{
Charles University, Faculty of Mathematics and Physics, Ke Karlovu 3, 121 16 Prague 2, Czech Republic
}

\date{\today}

\begin{abstract}

\setlength\intextsep{0pt}
\begin{wrapfigure}{r}{0.475\textwidth}
  \hspace{-1.5cm}
  \includegraphics{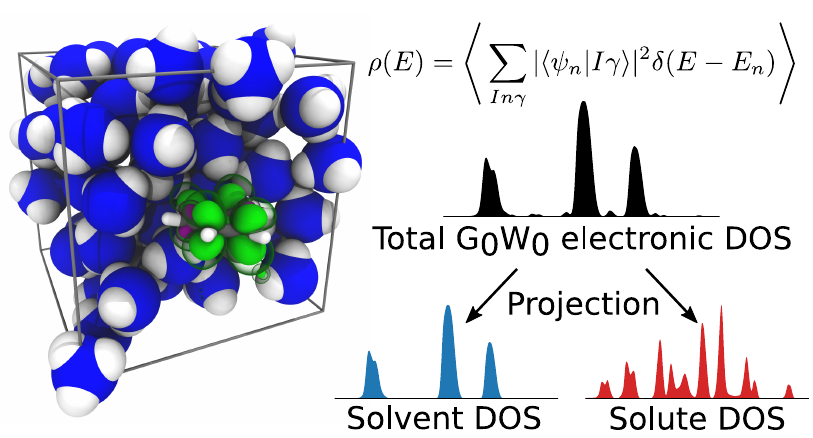}
\end{wrapfigure}

The benzene radical anion is a molecular ion pertinent to several organic reactions, including the Birch reduction of benzene in liquid ammonia.
The species exhibits a dynamic Jahn--Teller effect due to its open-shell nature and undergoes pseudorotation of its geometry.
Here we characterize the complex electronic structure of this condensed-phase system based on \textit{ab initio} molecular dynamics simulations and GW calculations of the benzene radical anion solvated in liquid ammonia.
Using detailed analysis of molecular and electronic structure, we find that the spatial character of the excess electron of the solvated radical anion follows the underlying Jahn--Teller distortions of the molecular geometry.
We decompose the electronic density of states to isolate the contribution of the solute and to examine the response of the solvent to its presence.
Our findings
show the correspondence between instantaneous molecular structure and spin density,
provide important insights into the electronic stability of the species, revealing that it is indeed a bound state in the condensed phase,
and offer electronic densities of states that aid in the interpretation of experimental photoelectron spectra.

\end{abstract}

{\maketitle}

\begin{bibunit}

\section{Introduction}

Liquid ammonia is particularly well-known as a solvent which sustains long-lived solvated electrons formed by the dissolution of alkali metals~\cite{Zurek2009/10.1002/anie.200900373}.
Recently, we used the flexible combination of refrigerated liquid microjet X-ray photoelectron spectroscopy~\cite{Faubel1997/10.1063/1.474034,Buttersack2019/10.1063/1.5141359} (XPS) and advanced \textit{ab initio} calculations to characterize the electronic structure of neat liquid ammonia~\cite{Buttersack2019/10.1021/jacs.8b10942} as well as the alkali metal solutions~\cite{Buttersack2020/10.1126/science.aaz7607}. 
In the latter, the hallmark XPS feature of the solvated electron is located at the electron binding energy of $-$2.0~eV relative to the vacuum level and its concentration dependence was used to experimentally map the electrolyte-to-metal transition~\cite{Buttersack2020/10.1126/science.aaz7607}.
Solvated electrons, essentially localized electrons bound in cavities formed within the solvent structure, act as powerful chemical reducing agents and, as such, find applications in numerous organic reductions. 
Arguably, the best known example is the Birch reduction of benzene in the environment of solvated electrons with the addition of an aliphatic alcohol~\cite{Birch1946/10.1038/158585c0}.
During the course of the reaction, the solvated electron binds to the benzene molecule, forming the benzene radical anion as the first reactive intermediate.
This chemical role of the benzene radical anion as well as its prominent position as the simplest example of an aromatic anion has prompted several experimental~\cite{Shida1973/10.1021/ja00792a005,Tuttle1958/10.1021/ja01553a005,Moore1981/10.1021/j150604a010,Sanche1973/10.1063/1.1679228} and theoretical~\cite{Hinde1978/10.1021/ja00483a010,Bazante2015/10.1063/1.4921261} studies of the species in the past.
A particularly intriguing conclusion that arises from these studies is that the stability of the species is environment-dependent.
In particular, the isolated benzene radical anion represents an unbound metastable shape resonance with a life time on the femtosecond time scale, which was consistently demonstrated both by \textit{ab initio} calculations~\cite{Hinde1978/10.1021/ja00483a010, Bazante2015/10.1063/1.4921261} and by electron scattering experiments~\cite{Sanche1973/10.1063/1.1679228} in the gas phase.
In contrast, the feasibility of the Birch reduction and various spectroscopic experiments performed in different polar solvents~\cite{Tuttle1958/10.1021/ja01553a005,Shida1973/10.1021/ja00792a005,Moore1981/10.1021/j150604a010}, which measure the species over extended time scales, imply the stability of the electronic structure of the benzene radical anion as well as its thermodynamic stability in the context of a chemical equilibrium with solvated electrons~\cite{Marasas2003/10.1021/jp026893u}. 
In addition to the non-trivial behavior of the electronic structure with respect to solvation, the presence of an excess electron in an initially energetically degenerate quantum state gives rise to a dynamic multimode $E \otimes e$ Jahn--Teller (JT) effect~\cite{OBrien1993/10.1119/1.17197,Bersuker2006} which results in complex behavior of the electronic structure as well as the molecular geometry.
In particular, the optimal molecular structure of the benzene radical anion is not a highly symmetric hexagonal one like that of the neutral benzene parent molecule, but is rather represented by a continuum of lower-symmetry structures that form the so-called pseudorotation path~\cite{Bazante2015/10.1063/1.4921261}.

Anticipating a future XPS measurement of the benzene radical anion as a natural continuation of the metal-ammonia solutions research, we have previously investigated the benzene radical anion in a liquid ammonia solution using computational methods with the aim to shed light on its structure, dynamics, and spectroscopy and to provide a theoretical basis to aid the interpretation of various experimental data.
In our original work~\cite{Brezina2020/10.1021/acs.jpclett.0c01505}, we performed \textit{ab initio} molecular dynamics (AIMD) of the explicitly solvated anion under periodic boundary conditions.
These simulations were realized at the hybrid density functional theory (DFT) level of electronic structure which we have shown to be, despite its high computational cost, a necessary methodological component to obtain a physically meaningful description of the benzene radical anion.
At this level of theory, the excess electron spontaneously localizes on the benzene ring and remains stable for the length of the simulation, indicating the presence of a bound electronic state.
Based on these simulations, we then addressed the structure of the solute and tracked the systematic geometry distortions and pseudorotation due to the dynamic JT effect that persist in the thermalized bulk system.
More recently, we approached the problem of the solvent-induced stability of the benzene radical anion from the point of view of molecular clusters derived from the original condensed-phase AIMD simulations~\cite{Kostal2021/10.1021/acs.jpca.1c04594}.
In that study, we calculated the excess electron vertical binding energy using explicit ionization in clusters of increasing size and found results ranging from $-$2.0 to $-$3.0~eV at the infinite cluster size limit depending on the specific methodology.

The present work aims to shed light on the electronic structure of the benzene radical anion by employing advanced electronic structure calculations and analysis performed on our original AIMD thermal geometries.
In the spatial domain, we describe the probability distribution of the excess electron and its correlation with the underlying JT distortions of molecular geometry using unsupervised machine learning methods~\cite{Glielmo2021/10.1021/acs.chemrev.0c01195}.
These methods have been used to analyze molecular dynamics trajectories and characterize representative molecular configurations of the studied systems in a bias-free way~\cite{Gasparotto2014/10.1063/1.4900655,Cerriotti2011/10.1073/pnas.1108486108}.
Here, we employ clustering analysis not only to the distribution of nuclear configurations of the benzene radical anion, but we also use it in conjunction with dimensionality reduction to characterize the electronic structure.
Then, in the energy domain, we aim to predict the binding energies of all the valence electrons in the studied system which can be directly compared to XPS data. 
To avoid the unphysical orbital energies directly available from the AIMD on-the-fly Kohn--Sham (KS) DFT electronic structure, we perform computationally demanding condensed-phase G$_0$W$_0$ calculations~\cite{Huser2013/10.1103/PhysRevB.87.235132,Wilhelm2016/10.1021/acs.jctc.6b00380} on the AIMD geometries to predict the electronic densities of states (EDOS).
To better understand the contributions of the individual species in the system in question, we additionally employ an approach which projects the EDOS on local atomic orbitals to resolve the calculated data by species and in space.

The rest of this paper is organized as follows.
In Section~\ref{sec:methodology}, we discuss the details of the performed simulations and calculations and describe the technical foundations of the employed analysis.
The main findings are then presented and discussed in Section~\ref{sec:results}.
There, we first focus on the results pertaining to the JT effect on the electronic structure and its correlation with the underlying molecular geometry.
Then, we move on to the energetics of the electronic structure and the question of the stability and binding energy of the solvated benzene radical anion.
These results are referenced against those for neutral benzene solvated in liquid ammonia and neat liquid ammonia itself.
Finally, we summarize our results and draw conclusions in Section~\ref{sec:conclusions}.

\section{Methodology}
\label{sec:methodology}

\subsection{AIMD simulations}

The original AIMD simulations of the benzene radical anion and neutral benzene in liquid ammonia under periodic boundary conditions were realized using the the CP2K 5.1 package~\cite{Hutter2014/10.1002/wcms.1159, Guidon2008/10.1063/1.2931945,Guidon2009/10.1021/ct900494g} and its Gaussian and plane wave~\cite{Lippert1997/10.1080/002689797170220}
electronic structure module Quickstep~\cite{Vandevondele2005/10.1016/j.cpc.2004.12.014}.
Both simulated systems consisted of one solute molecule and 64 solvent molecules in a cubic box of a fixed side length of 13.745~\AA\ and 13.855~\AA\ for the benzene radical anion and neutral benzene, respectively.
The nuclei were propagated with a 0.5~fs time step in the canonical ensemble at 223~K using the stochastic velocity-rescaling thermostat~\cite{Bussi2007/10.1063/1.2408420}.
The electronic structure was calculated using the revPBE0-D3 hybrid density functional~\cite{Perdew1996/10.1103/PhysRevLett.77.3865, Zhang1998/10.1103/PhysRevLett.80.890, Adamo1999/10.1063/1.478522, Goerigk2011/10.1039/c0cp02984j} to limit the self-interaction error, as required for the localization of the excess electron and the stability of the benzene radical anion~\cite{Brezina2020/10.1021/acs.jpclett.0c01505}.
The KS wavefunctions were expanded into the TZV2P primary basis set~\cite{VandeVondele2007/10.1063/1.2770708}, while the density was expanded in an auxiliary plane-wave basis with a 400~Ry cutoff.
GTH pseudopotentials~\cite{Goedecker1996/10.1103/PhysRevB.54.1703} were used to represent the core $1s$ electrons of the heavy atoms.
Additionally, the auxiliary density matrix method~\cite{Guidon2010/10.1021/ct1002225} with the cpFIT3 auxiliary basis set~\cite{Guidon2010/10.1021/ct1002225} was used to accelerate the computationally demanding hybrid DFT electronic structure calculations.
The total simulated time was 100~ps for both systems, each collected from five 20~ps trajectories initialized from decorrelated and equilibrated initial conditions.

\subsection{G$_0$W$_0$ calculations}

In this work, we use the G$_0$W$_0$ method~\cite{Huser2013/10.1103/PhysRevB.87.235132,Wilhelm2016/10.1021/acs.jctc.6b00380} which gives access to physically meaningful one-electron energy levels of the investigated condensed-phase, periodic systems.
This is in contrast with orbital energies of the underlying KS DFT which should not formally be considered as one-electron energies.
For each solute, these calculations are performed on top of 205 DFT-AIMD thermal structures extracted from the AIMD trajectories with a 0.5~ps stride with the revPBE0-D3/TZV2P KS wavefunctions being used as a starting point to obtain the corrected G$_0$W$_0$ energies.
These calculations are realized using the CP2K package, version 7.1.
The self-energy is described analytically over the real frequency axis using the Padé approximation and the Newton-Raphson fixed point iteration is employed for numerical solution of the corresponding algebraic equations.
The influence of periodic boundary conditions on the G$_0$W$_0$ energies is minimized by employing a periodicity correction scheme~\cite{Wilhelm2017/10.1103/PhysRevB.95.235123}.
The resulting EDOS, obtained as the distribution of the G$_0$W$_0$ energies, is described as a continuous probability density function through the kernel density estimation method using a Gaussian kernel with a 0.02~eV bandwidth.
The G$_0$W$_0$ calculations performed in periodic boundary conditions do not directly provide the absolute values of electron binding energies due to the absence of the explicit liquid-vacuum boundary.
Thus, to access the absolutely positioned EDOS, the whole spectrum must be shifted on the energy axis by a suitable constant.
In other works, this was achieved by auxiliary slab calculations that provide an estimate of the shift~\cite{Ambrosio2018/10.1021/acs.jpclett.8b00891,Gaiduk2018/10.1038/s41467-017-02673-z}.
In our previous work on neat liquid ammonia combining G$_0$W$_0$ calculations with liquid XPS~\cite{Buttersack2019/10.1021/jacs.8b10942,Faubel1997/10.1063/1.474034}, we aligned the average energy of the calculated liquid $\mathrm{3a_1}$ peak to $-$9.09~eV, the average of the same peak obtained experimentally.
This bypassed the need for additional \textit{ab initio} calculations and facilitated the comparison of the whole spectrum between theory and experiment.
Here, we exploit the fact that, as detailed in Section~\ref{sec:results}, the electronic perturbation of the liquid ammonia solvent by the presence of the benzene radical anion is minor.
As such, the total EDOS was shifted to match the same experimental valence liquid ammonia peak as in our previous work.
The value of the shift was determined from the mean energy of the $\mathrm{3a_1}$ ammonia peak of the total EDOS with the solute included (other options are discussed in Section~\ref{sec:additional-results} of the supplementary material).

To gain insight into the contributions of individual chemical species to the total G$_0$W$_0$ EDOS, we decompose this quantity into separate densities for each species and address the differences between the neat ammonia data and the data from systems with solutes.
Specifically, we rely on the original formulation of the projected density of states (PDOS) for KS orbitals~\cite{Hunt2003/10.1016/S0009-2614(03)00954-0}, which projects the total EDOS on the respective part of the atomic orbital basis set of every atom in the system individually.
Extending the original approach, we use these projections for the G$_0$W$_0$-corrected binding energies, since the spatial orbitals are identical between KS DFT and G$_0$W$_0$.
For each atom and each configuration, each G$_0$W$_0$ energy is assigned a weight based on the magnitude of the projection of the corresponding orbital on that atom.
Naturally, these atomic contributions can be collected into molecular contributions as needed for each particular system.
The total EDOS can be expressed as the following ensemble average over the contributing structures
\begin{equation}
    \rho(E) = \left\langle \sum_n \delta(E - E_n) \right\rangle,
\end{equation}
where $E_n$ are the G$_0$W$_0$ one-electron energy eigenvalues and angle brackets denote an average over the ensemble of thermal structures.
To decompose it, we use a projection on an atom-centered linear combination of atomic orbitals (LCAO) basis set $\{\ket{I\gamma}\}$.
This basis satisfies the completeness relation over the spanned space
\begin{equation}
    \sum_I \sum_{\gamma} \ket{I\gamma}\bra{I\gamma} = \hat{\mathbf{1}},
\end{equation}
where the summation runs over all atoms $I$ and all additional quantum numbers $\gamma$ and $\hat{\mathbf{1}}$ denotes the identity operator.
Using the orthonormality of the original KS orbitals $\ket{\psi_n}$ that remain unchanged during the G$_0$W$_0$ calculation, we can expand the total EDOS definition as a sum over atomic projections as
\begin{equation}
\begin{split}
    \rho(E)
    & = \left\langle \sum_n \braket{\psi_n}{\psi_n} \delta(E - E_n) \right\rangle \\
    & = \left\langle \sum_n \sum_I \sum_\gamma \braket{\psi_n}{I\gamma}\braket{I\gamma}{\psi_n} \delta(E - E_n) \right\rangle \\
    & = \sum_I \left\langle \sum_n \sum_\gamma |\braket{\psi_n}{I\gamma}|^2 \delta(E - E_n) \right\rangle \\
    & \equiv \sum_I \langle S_I(E) \rangle 
    \equiv \sum_I \rho_I (E),
\end{split}
\end{equation}
where we have labeled the overlap-weighted kernel of the thermal average $S_I(E)$ and the whole thermally averaged atomic projection $\rho_I (E)$.
These atomic projections can then be summed over arbitrary subsets of atoms to obtain a PDOS on any species in question.
Moreover, we can further resolve the atomic contributions as a function of distance $r$ from a chosen point of reference
as the following two dimensional distribution
\begin{equation}
    \rho(E, r)
    =
    \frac{1}{4\pi r^2 g(r)}
    \sum_I \langle S_I(E) \delta(r - r_I) \rangle,
\end{equation}
where $r_I$ is the distance of the $I$-th atom from the point of reference and the normalization factor in the denominator based on the radial distribution function $g(r)$ of the chosen species around the same point of reference ensures a uniform marginal distribution in $r$.

\subsection{Clustering Analysis}

The clustering of the relevant feature space vectors is based on the Gaussian Mixture Model (GMM) as implemented in the scikit-learn Python library~\cite{Pedregosa2012/}.
For the molecular geometries, we assign features as all vibrational normal modes with JT-active symmetry to naturally describe the distortions in an 8D configuration space.
For the electronic structure, we use a high-dimensional abstract feature space that relies on a Fourier decomposition of the respective spin densities.
Both feature spaces are described in detail in the following paragraphs.
The GMM algorithm was chosen over the commonly used $k$-means clustering since it allows to reach a similar goal in a more flexible and general way and, moreover, yields a continuous parametrization of the obtained clusters in terms of high-dimensional Gaussian functions that can be used to evaluate the cluster membership probability.
The full covariance in all dimensions was employed to account for possible spatial anisotropy of the clusters and a tight convergence limit of $10^{-5}$ was used.

\section{Results and Discussion}
\label{sec:results}

In the following paragraphs, we focus on the spatial character of the excess electron of the benzene radical anion using the spin density, an observable quantity obtained directly from an unrestricted Kohn--Sham (KS) DFT calculation.
We aim at a description of the evolution of the spin density in the context of the condensed-phase JT effect, which governs the distortions of the underlying molecular geometry of the benzene radical anion solvated in liquid ammonia~\cite{Brezina2020/10.1021/acs.jpclett.0c01505}.
Specifically, we ask if the molecular distortions correlate with the immediate shape of the spin density and, therefore, if information about the JT state of the solute can be extracted directly from the electronic structure of the solvated species, similarly to how it can be extracted from its molecular geometry.
Later on, we turn our attention to the energetics of the electronic structure, predict electronic densities of states for the studied system, and discuss then in detail in the context of the question of the stability of the solvated benzene radical anion as well as from the perspective of interpretation of XPS data.

\subsection{The Jahn--Teller Effect on the Molecular and Electronic Structure}

The essence of the JT effect in the benzene radical anion is as follows.
As the $D_{6h}$-symmetric benzene molecule accepts an excess electron, the formed degenerate $\mathrm{E_{2u}}$ electronic state of the radical anion becomes unstable since it corresponds to a conical intersection between two adiabatic potential energy hypersurfaces (APESs).
This instability is resolved by a symmetry-lowering distortion along the JT-active normal modes of $\mathrm{e_{2g}}$ symmetry, which brings the system into a minimum on the pseudorotational path on the lower branch of the JT-split APES.
At the same time, the symmetry of the initial electronic state is reduced as well,
with two new possible lower symmetries, $\mathrm{A_u}$ and $\mathrm{B_{1u}}$, corresponding to the ground state of the benzene radical anion in the two opposite distortions of the molecular geometry.

\subsubsection{Clustering of Molecular Geometries}

The natural coordinates to describe the molecular distortions are the four degenerate pairs of JT-active normal modes.
These are adopted here consistently with our previous work from the vibrational normal modes of an optimized neutral benzene molecule since it shares the same molecular structure and the point group with the radical anion in its reference undistorted geometry.
A physically meaningful observation of the JT pseudorotation can be made by averaging the full 8D data over all modes that do not exhibit a strong enough JT split to be observable in the thermal system.
Thus, the pseudorotation can be represented as a 2D distribution in the pair of remaining $\mathrm{e_{2g}}$ modes at 1654~cm$^{-1}$ which show an appreciably strong JT effect.
In this case, the free energy landscape of the pseudorotation valley is essentially flat and the path around it is described by the pseudorotation angle $\theta =\arctan2(Q_y / Q_x)$, a scalar parameter which represents the polar angle in the 2D subspace of the relevant normal mode coordinates labeled as $Q_x$ and $Q_y$~\cite{Brezina2020/10.1021/acs.jpclett.0c01505}.

In order to analyze the full 8D distribution, we applied the GMM clustering algorithm to the normal modes data set with the aim to find representative distortions.
However, unlike in the case of the electronic structure discussed in the following paragraphs, the resulting clustering of the data is not satisfactory for several reasons.
Motivated by the threefold symmetry of the reference gas-phase APES, we attempted to separate the data into both three and six clusters.
In both cases, clustering of comparable quality was obtained which implies that there is no clear number of natural clusters in the data set.
This is further supported by additional attempts to cluster the data into a number of clusters that does not respect the inherent symmetry of the problem: again, similar outputs were produced.
Moreover, the clustering is generally not reproducible and inconsistent positions of clusters are obtained each time.
As a measure of clustering performance, we use silhouette coefficients, which range from $-$1 (wrong clustering) through 0 (poor clustering) to $+$1 (excellent clustering)~\cite{Rousseeuw1987/10.1016/0377-0427(87)90125-7}.
If we cluster our data into three groups, the average silhouette coefficient does not exceed the value of $\sim$0.08, which quantifies the insufficient separation of the data (a silhouette plot is presented in Section~\ref{sec:additional-results} of the supplementary material).
This demonstrated lack of clear separation in the molecular geometries suggests that the remaining modes do not bring much additional structure to the data set in comparison to the reduced 2D distribution in $Q_x$ and $Q_y$ and the essentially flat character of the probability distribution around the pseudorotation valley generalizes to the full dimensionality.
Therefore, we adhere to the simpler continuous parametrization by the pseudorotation angle $\theta$ to describe molecular distortions in the following analysis.

\subsubsection{Spin density dimensionality reduction}

\begin{figure}[tb!]
    \centering
    \includegraphics[width=\linewidth]{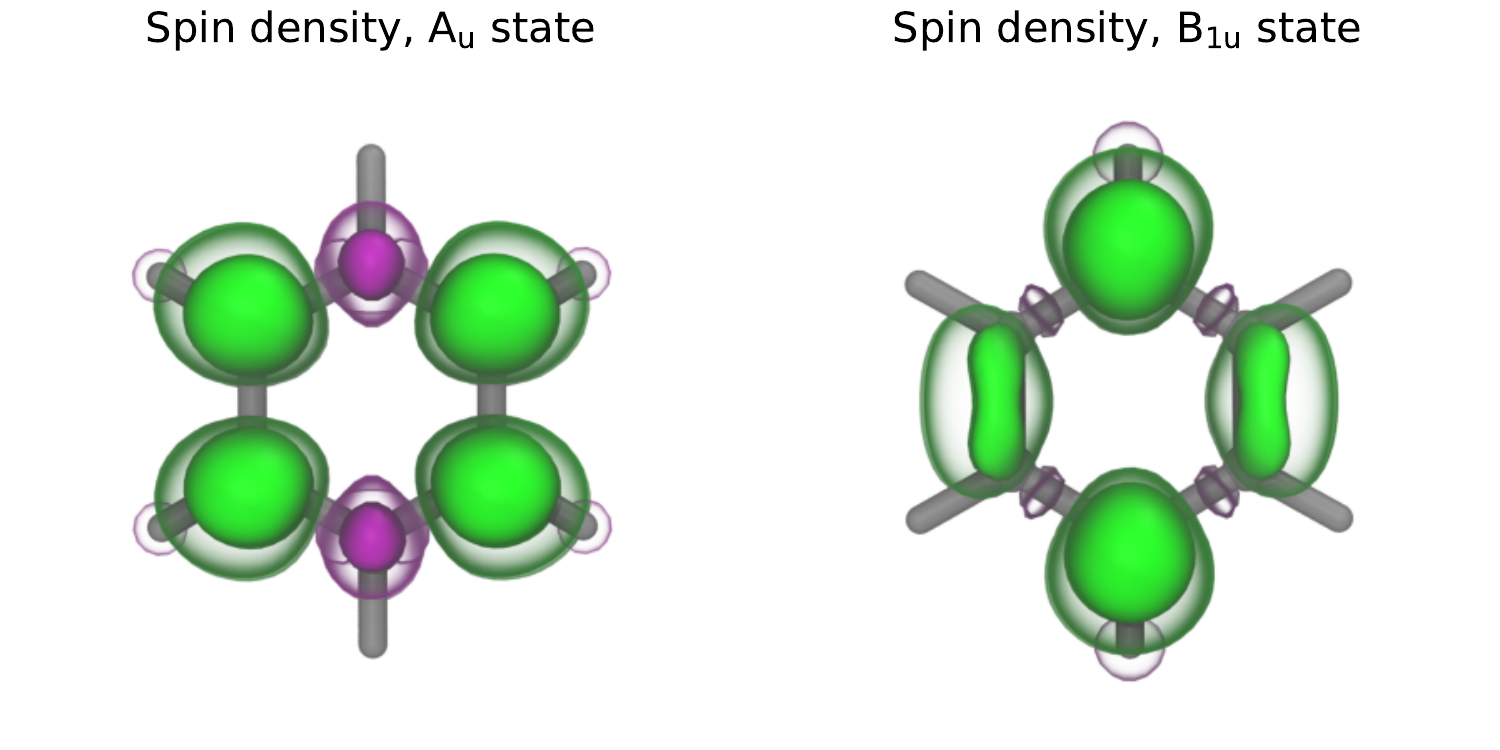}
    \caption{The spin densities of the $\mathrm{A_{u}}$ and $\mathrm{B_{1u}}$ electronic states of the benzene radical anion.
    The presented idealized geometries and spin densities were obtained from a finite basis set gas-phase calculation at the hybrid DFT level, as used for the AIMD simulations; similar spin densities are however observed in the condensed-phase simulations.
    The positive deviations of the spin density are shown in green at two contours, 0.025~$a_0^{-3}$ (opaque), and 0.006~$a_0^{-3}$ (transparent), while the negative deviations are shown in purple at the same isovalues with a negative sign.
    The molecular structure of the benzene radical anion is shown in gray as a whole.}
    \label{fig:spin_density}
\end{figure}

To motivate the analysis of the electronic structure of the solvated species, we consider optimized gas-phase benzene radical anion structures where the excess electron is artificially localized due to a finite orbital basis set.
The spin density distributions for the two distinct JT distoritions are shown in Figure~\ref{fig:spin_density}.
The spin density of the $\mathrm{A_u}$ state (left) is characterized by four atom-centered maxima and two less pronounced minima localized along one of the $C_2$ symmetry axes; the $\mathrm{B_{1u}}$ spin density (right) exhibits two maxima localized on distal carbon atoms along the corresponding $C_2$ axis and two elongated bridge-like positive deviations over a pair of carbon--carbon bonds parallel with this $C_2$ axis.
We also have to take into account that the high symmetry of the benzene molecular geometry allows for distortion in three equivalent directions corresponding to the three horizontal, apex-to-apex $C_2$ axes in the $D_{6h}$ point group.
These distortions are represented by two sets of three equivalent stationary points around the point of high symmetry on the pseudorotation APES, each separated by a pseudorotation angle of 60$^\circ$ from its opposite-kind neighbors and by 120$^\circ$ from its pseudorotated images.
As a consequence, three equivalent $\mathrm{A_u}$-type and three equivalent $\mathrm{B_{1u}}$-type spin densities exist which correspond to the six APES stationary points.
The pseudorotation of the nuclear geometry between these minima is discussed in detail in Reference~\citenum{Brezina2020/10.1021/acs.jpclett.0c01505}; a video file illustrating the evolution of the spin density on top of the pseudorotating geometry in the idealized gas-phase case is included in the supplementary material and described in Section~\ref{sec:SI_videos}.

\begin{figure}[tb!]
    \centering
    \includegraphics[width=\linewidth]{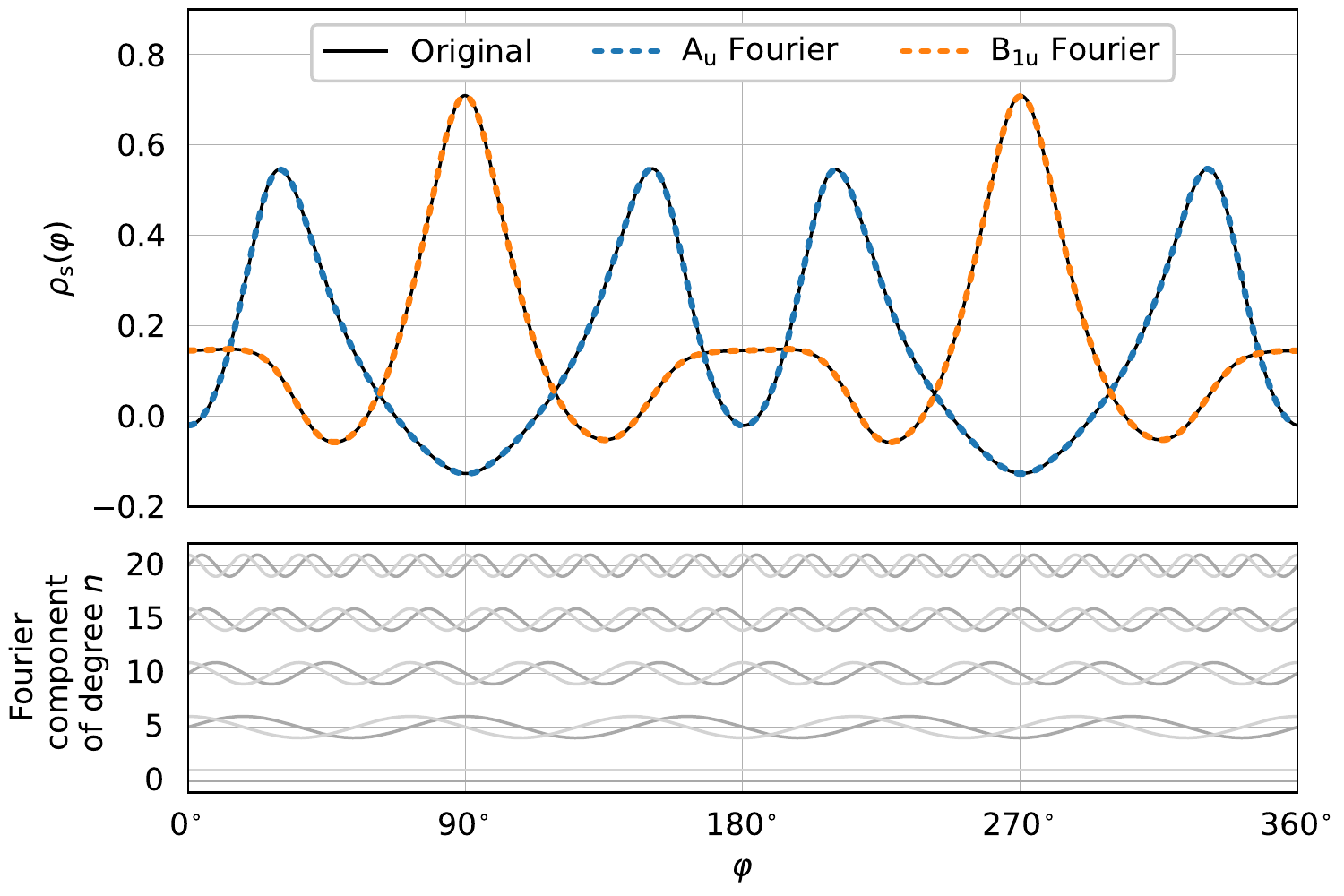}
    \caption{The relevant 1D spin densities $\rho_\mathrm{s}(\varphi)$ for the $\mathrm{A_u}$ and $\mathrm{B_{1u}}$ states (Figure~\ref{fig:spin_density}).
    The original functions are shown as black solid lines.
    The $N = 20$ Fourier-reconstructed curves are shown as dashed blue and orange lines, respectively.
    The bottom panel illustrates five samples equidistant in their degree $n$ from the employed $N = 20$ Fourier basis with the sine components shown in dark gray and cosine components in light gray.}
    \label{fig:1D-spin_density}
\end{figure}

Eventually, we want to analyze the thermal data in the condensed phase---the natural environment where the solvated benzene radical anion is electronically stable and physically relevant observation of the JT effect and the associated spin density can be made.
For this purpose, we design a two-step dimensionality reduction procedure that represents the spin densities in a feature space of reasonable dimension in such a way that the two idealized JT-distorted cases can be distinguished.
The periodicity of the spin density along the aromatic ring
makes it advantageous to express its spatial dependence in terms of a local spherical coordinate system $r, \vartheta$ and $\varphi$ (where $\varphi$ is the polar angle ranging from 0 to $2\pi$).
These coordinates are obtained by the usual transformation from a local Cartesian system 
in which the $x, y$-plane is represented by the molecular plane of the benzene radical anion and the $z$-axis by its normal with its origin at the solute center of mass (see Section~\ref{sec:analysis-details} of the supplementary material for details).
The spherical coordinates represent a natural description for the systems in question and allow to reduce the dimensionality of the full spin density into a one-dimensional (1D) function by partial integration.
As documented in Section~\ref{sec:additional-results} of the supplementary material, the 1D spin densities in $r$ and $\vartheta$ show practically perfect overlap for the two spin density types and thus bring no distinction between them.
The information that distinguishes the two types is contained in the remaining possible spin density in $\varphi$
\begin{equation}
    \rho_\mathrm{s}(\varphi) =
    \int_{0}^{\pi} \int_{0}^{r_{\mathrm{max}}} \dd \vartheta \dd r \ r^2 \sin \vartheta \rho_\mathrm{s}(r, \vartheta, \varphi),
\end{equation}
that describes the character of the spin density around the benzene ring. 
Its shape can be traced back to the spatial characteristics of the full spin densities through the respective sequence of the 1D maxima and minima along the aromatic ring, as shown for the idealized spin densities in Figure~\ref{fig:1D-spin_density}, top panel, full lines.
In terms of $\rho_\mathrm{s}(\varphi)$, the pseudorotation of each type of the full spin density by 120$^\circ$ translates simply into a 120$^\circ$ shift on the $\varphi$-axis.

At this point, the 3D spin density is reduced to a 1D function that is still fully capable of distinguishing between the two spin density types.
An additional level of simplification that opens the door to numerical analysis is achieved by mapping the continuous $2\pi$-periodic 1D spin densities onto discreet vectors by means of a Fourier series and noting that only the first few harmonics are necessary to achieve a highly accurate decomposition as demonstrated by the dashed curves in the top panel of Figure~\ref{fig:1D-spin_density}.
This set of Fourier coefficients clearly distinguishes the two idealized spin densities in relatively few dimensions.
While the technical aspects of this step are discussed in detail in Section~\ref{sec:analysis-details} of the supplementary material, we note here that the Fourier decomposition was performed using the first 20 harmonics, yielding an 82-dimensional Euclidean feature vector for each spin density sample (a total of $2(2N+1)$ real coefficients are needed for a Fourier series counting $N$ harmonic functions).

\subsubsection{Clustering of the Electronic Structure}

We are now able to represent each spin density distribution in a compact way and can move to the analysis of the electronic structure of the condensed-phase system.
A visual inspection of the trajectory~\cite{Brezina2020/10.1021/acs.jpclett.0c01505} of the solvated benzene radical anion clearly reveals the presence of two limiting spin density structures similar to the optimized ones.
Therefore, we aim to perform an analysis that will allow us to divide the observed ensemble of condensed-phase spin densities into six categories centered around each of the limiting spin density structures and including the surrounding thermal population.
Once this is established, one can examine the correlation between the immediate electronic structure and the underlying molecular geometry of the solute.

\begin{figure}[tb!]
    \centering
    \includegraphics[width=\linewidth]{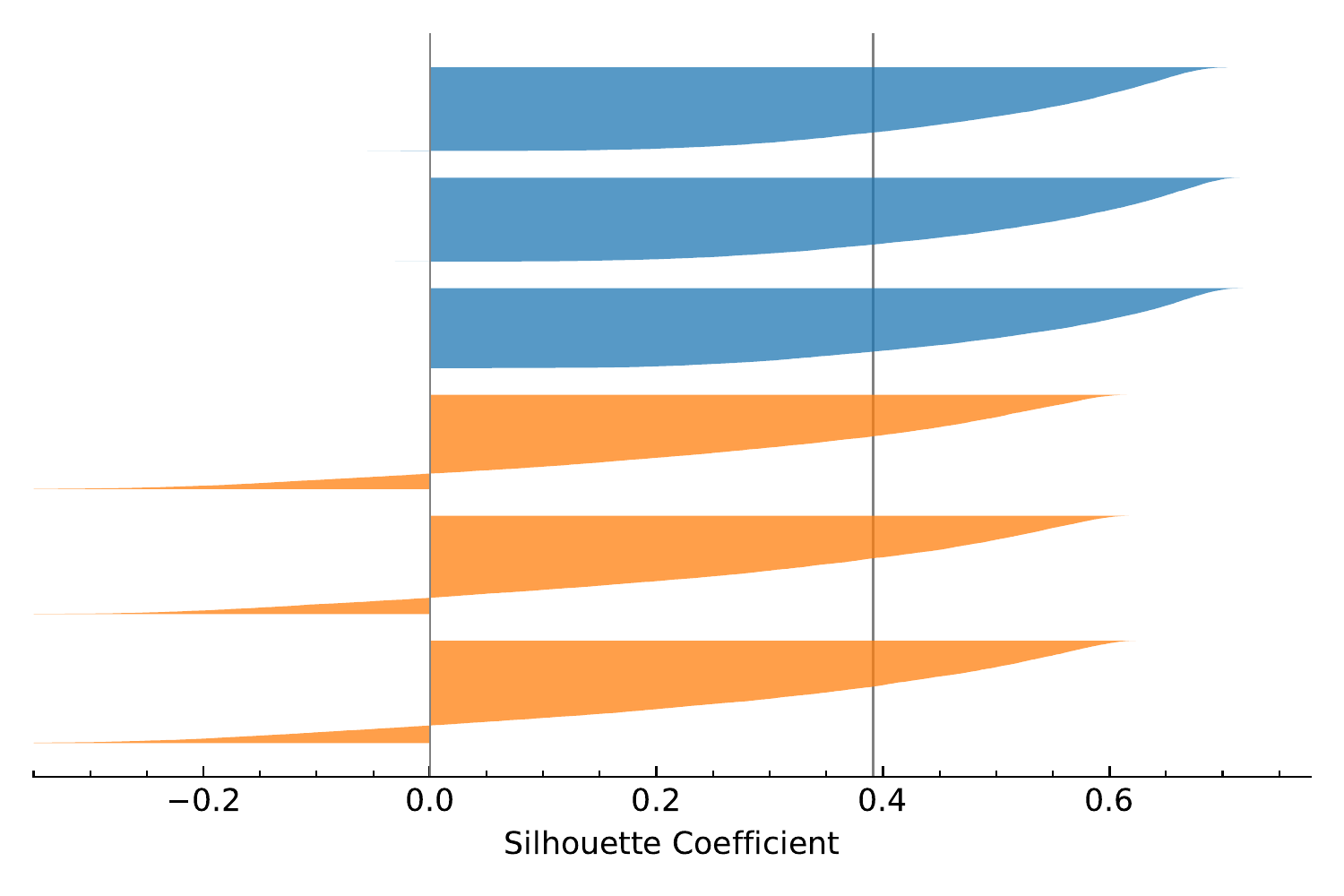}
    \caption{Characterization of the clustering of the electronic structure by the means of silhouette plots.
    The top three clusters represent the $\mathrm{A_u}$ clusters, the bottom three the $\mathrm{B_{1u}}$ clusters.}
    \label{fig:Silhouette-plot}
\end{figure}

To categorize the spin densities of the thermal solvated system, we turn again to GMM clustering to separate the data now concisely represented as feature vectors constructed out of Fourier coefficients.
GMM is able not only to split the data into natural clusters, but also to provide a continuous parametrization of each cluster through evaluation of posterior probabilities of cluster membership.
Indeed, in this case, the data set splits cleanly into six clusters as shown by the cluster silhouettes presented in Figure~\ref{fig:Silhouette-plot} which average to the mean silhouette coefficient of $\sim$0.4 and contain no outliers for the $\mathrm{A_u}$ state and only a small number of outliers (negative silhouette coefficients) for the $\mathrm{B_{1u}}$ state.
Additional clustering validation is documented in Section~\ref{sec:additional-results} of the supplementary material.
The centers of the six clusters then correspond to the electronic structure at the six possible $D_{2h}$ distortions and the population of each cluster corresponds to the thermal fluctuations around these minima.
This is directly shown by summing up the Fourier series defined by the coordinates of the cluster centers to obtain new 1D spin densities.
These exhibit physically meaningful properties such as close-to-reference shapes (such as those shown in Figure~\ref{fig:1D-spin_density}) and the expected 120$^\circ$ shifts within each type group (see the supplementary material, Section~\ref{sec:additional-results}).
While these findings show that the excess electron structure is analogous to that found for the benzene radical anion in the gas phase using a comparable finite orbital basis set, it is important to keep in mind that such a system converges to an unbound state when the size of the basis set is increased.
Only in the condensed phase is the species actually bound and its JT effect observable and the electronic states physically meaningful and potentially experimentally measurable.

\subsubsection{Correlation between the electronic structure and molecular geometry}

\begin{figure}[t!]
    \centering
    \includegraphics[width=\linewidth]{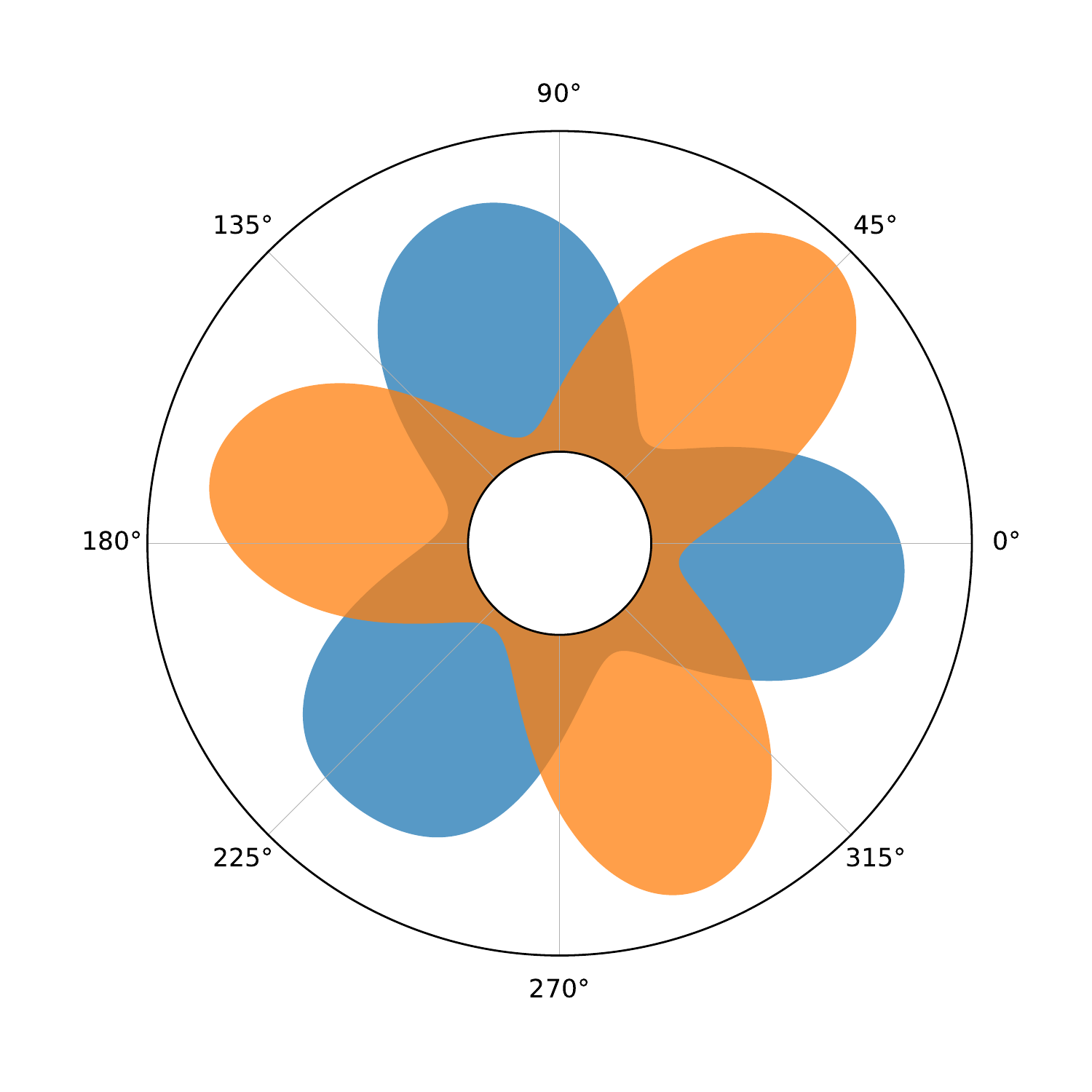}
    \caption{Correlation of the electronic structure with the distortion of nuclear geometry. 
    Each molecular distortion is characterized here by the value of the pseudorotation angle $\theta$.
    The distributions of $\theta$ weighted by the corresponding electronic parameters $p(\mathrm{A_u})$ (blue) and $p(\mathrm{B_{1u}})$ (orange) are shown in polar coordinates with an offset zero-distance.}
    \label{fig:JT}
\end{figure}

To quantify the correlation between the molecular structure and the spin density we exploit the features of the trained Gaussian mixture model to assign a posterior probability of belonging to a specific cluster to each spin density data point.
Thus, a generalized single-valued parameter $p(\mathrm{A_u})$, which can be defined as a sum over all $\mathrm{A_u}$-type cluster probabilities, gives the overall probability that a data point is of the $\mathrm{A_u}$-type, including all three possible pseudorotations. 
Clearly, the same can be done for the $\mathrm{B_{1u}}$-type clusters and the identity for complementary probabilities that $p(\mathrm{A_u}) + p(\mathrm{B_{1u}})= 1$ has to hold.
Now, since each spin density data point has a unique molecular geometry associated with it, the proposed probability parameters can be directly correlated with the underlying molecular distortions characterized by the pseudorotation angle $\theta$ as defined above.

We use these electronic probability parameters to weight each point contributing to the probability distribution in $\theta$, which is almost uniform originally.
This splits it into two distinct distributions, each with three well-defined peaks separated by a 120$^\circ$ increment.
These are shown in Figure~\ref{fig:JT}, exploiting a representation in polar coordinates with an offset origin.
The presented complementary distributions clearly show that the individual symmetries of the molecular distortions are accompanied by spin densities of the same type, as can be deduced from the fact that the distortion at $\theta = 0^\circ$ is uniquely identified with the distortion of molecular geometry corresponding to the $\mathrm{A_u}$ electronic state.
It thus appears that the electronic character of the JT effect of the benzene radical anion in liquid ammonia follows closely the predicted gas-phase theory while the solvent acts as a stabilizing, but non-perturbing environment.
Due to the correlation shown in Figure~\ref{fig:JT}, we conclude that similar information about the JT effect can be extracted from the immediate spin density as well as from the immediate molecular geometry of the solute.

Even though the molecular geometries undergo almost free pseudorotation with effectively no free energy barriers and can not therefore be clustered into distinct populations of different pseudorotamers, the situation is different for the electronic state of the system.
As it moves along the pseudorotation path, it transitions rather sharply between ground-state spin densities of the two possible symmetries, as revealed by our analysis.

\subsection{Energetics of the Electronic Structure}

\begin{figure}[tb!]
    \centering
    \includegraphics[width=\linewidth]{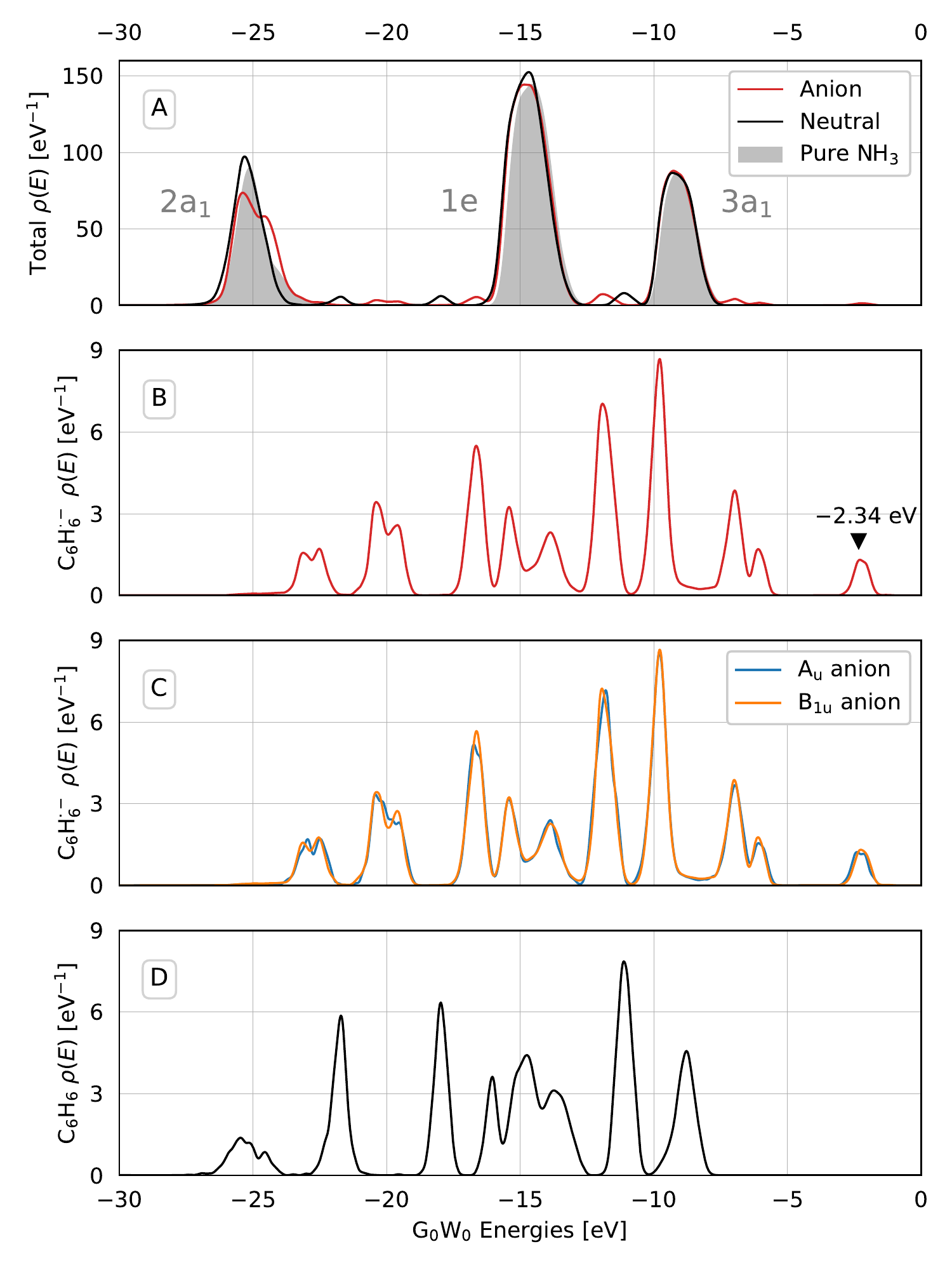}
    \caption{The solvated benzene radical and neutral benzene total G$_0$W$_0$ EDOS and the PDOS projections on the solutes. 
    Panel A: the total EDOS of the benzene radical anion (red) and neutral benzene (black) in liquid ammonia. 
    The calculated pure liquid ammonia EDOS~\cite{Buttersack2019/10.1021/jacs.8b10942} is shown in gray.
    Consistently with the published pure ammonia data, the corresponding peaks are labeled by the symmetry labels of the gas-phase ammonia molecular orbitals.
    Panel B: PDOS of the benzene radical anion.
    The projection shows a detailed account of the electronic structure of the anion, including the highest occupied state, marked by its binding energy and a black triangle.
    Panel C: The benzene radical anion PDOS resolved for the two type of JT-relevant electronic structure symmetries.
    Note that the small differences between the blue ($\mathrm{A_{u}}$) and orange ($\mathrm{B_{1u}}$) curves, caused by sampling from the corresponding smaller subsets of the calculated G$_0$W$_0$ energies, are insignificant within the available statistics.
    The PDOS of both JT structures is therefore identical.
    Panel D: PDOS of neutral benzene in liquid ammonia.}
    \label{fig:edos}
\end{figure}

At this point, we turn our attention to the energetics of the electronic structure of the whole studied system in terms of one-electron levels.
The single-electron energies are calculated using the G$_0$W$_0$ method~\cite{Huser2013/10.1103/PhysRevB.87.235132,Wilhelm2016/10.1021/acs.jctc.6b00380} on an ensemble of 205 structures drawn with a 0.5~ps stride from our previously published hybrid DFT trajectories of the benzene radical anion as well as neutral benzene for comparison.
The absolute energies of the whole spectrum were shifted as detailed in Section~\ref{sec:methodology}.
The distribution of the obtained G$_0$W$_0$ quasiparticle energies, which accurately approximate electron binding energies, represents the EDOS and is shown in panel A of Figure~\ref{fig:edos}.
The dominant three-peak pattern in both systems can be readily related to the neat liquid ammonia EDOS~\cite{Buttersack2019/10.1021/jacs.8b10942}, shown here in gray shading for reference.
In our systems with solutes, it is accompanied by a multitude of low-intensity features along the whole range of energies.
We can now use the projection approach detailed in Section~\ref{sec:methodology} to isolate these features and examine the solute and solvent spectra separately.

Focusing first on the benzene radical anion, we obtain the solute PDOS shown in panel B of Figure~\ref{fig:edos}.
Clearly, this component isolates the low-intensity features that do not overlap with the neat ammonia EDOS and, moreover, uncovers additional ones that were previously contained in the high-intensity solvent peaks.
Most notably, this solute PDOS suggests that the highest energy state, occupied by the excess electron, is fully accounted for by the solute, consistent with the previously observed spatial localization of the spin density~\cite{Brezina2020/10.1021/acs.jpclett.0c01505}.
Its mean binding energy of $-$2.34~eV and the absence of tails extending into the positive values prove that the excess electron on benzene is bound relative to the vacuum level, thus conclusively answering the question of stability of the molecular structure of the anion as long as it is solvated in liquid ammonia.
This is in excellent agreement with the vertical electron binding energy of $-$2.30~eV obtained by explicit ionization calculations of benzene radical anion and ammonia gas-phase clusters in the infinite cluster size limit~\cite{Kostal2021/10.1021/acs.jpca.1c04594}.  
Compared to neutral benzene (Figure~\ref{fig:edos}, panel D), the whole G$_0$W$_0$ anion solute PDOS is systematically shifted towards weaker binding energies by several electronvolts.
Its shape is modified as well, including several peak splittings not observed in the neutral system.
These are likely due to the overall lower symmetry of the anion, rather than due to the presence of two distinct JT pseudorotamers, which give rise to identical PDOS within the available statistical sampling, as shown in panel C of Figure~\ref{fig:edos}.
Since the excess electron binding energy in the benzene radical anion is close to the binding energy of the solvated electron of $-2.0$~eV~\cite{Buttersack2020/10.1126/science.aaz7607}, an overlap might arise in an experimental photoelectron spectrum if the two species coexist in equilibrium, leading to a single broader peak or perhaps a double peak feature.
This suggests that the excess electron binding energy itself might not be sufficient to prove the presence of the benzene radical anion.
However, a viable workaround exists in the predicted changes of the lower electronic levels of benzene after the addition of the excess electron.
These are large enough to be measured and several bands are localized in the regions where no overlap with the solvent signal is expected, as clearly shown by the projected densities.

\begin{figure}[tb!]
    \centering
    \includegraphics[width=\linewidth]{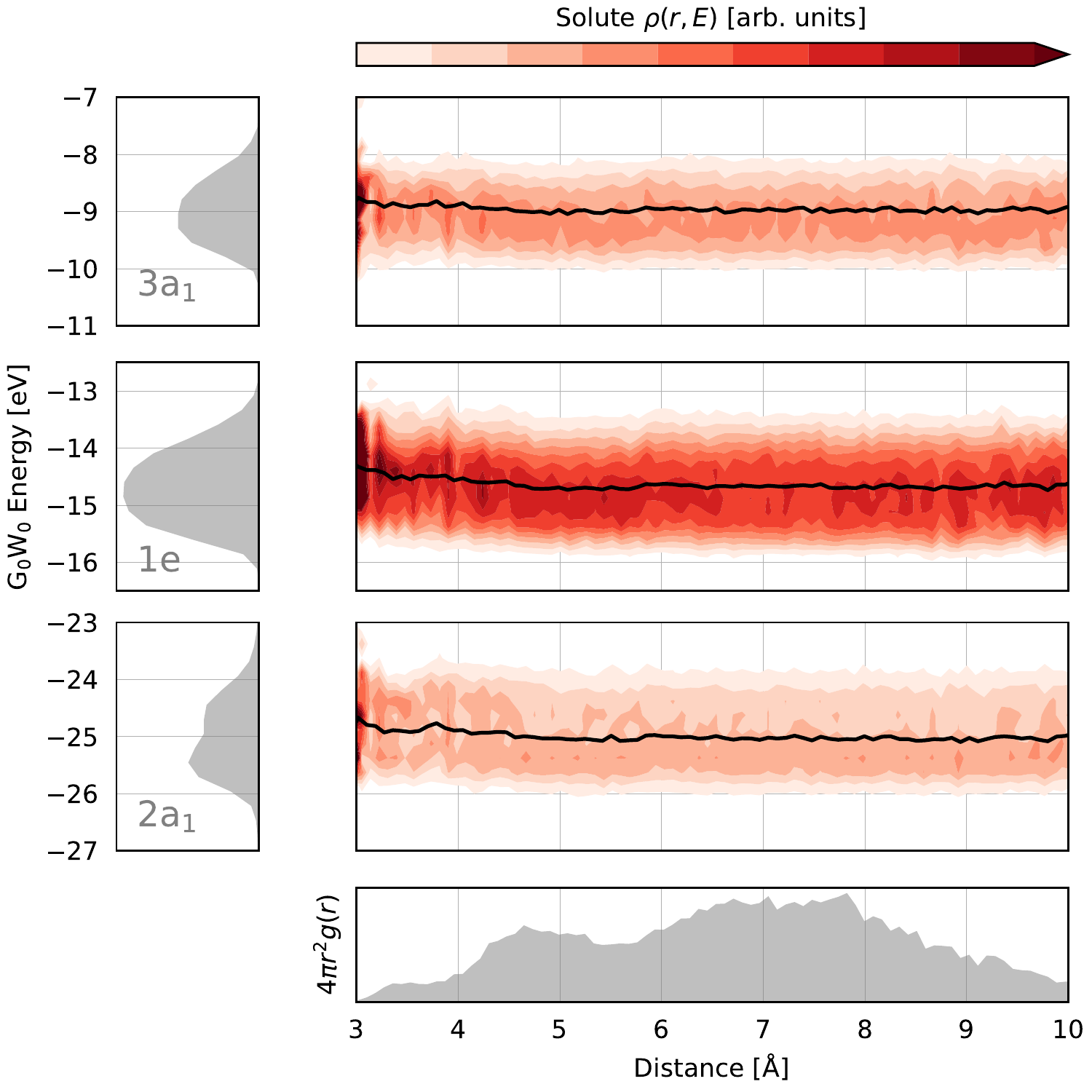}
    \caption{Electronic density of states projected on the solute subspace and resolved as a function of distance from the center of mass of the radical anion. 
    Black dashed lines denote the mean of each peak again as a function of distance.
    The left side panel shows the total solute PDOS in gray.}
    \label{fig:dr-pdos}
\end{figure}

Next, we concentrate on the solvent subspace.
In Figure~\ref{fig:dr-pdos}, the solvent PDOS shown in the left-hand side panels in gray shading features subtle differences compared to the EDOS of neat ammonia.
These appear because of the changes of the electronic structure of the solvent molecules induced by the interaction with the radical anion solute.
To better quantify this perturbation, we exploit the molecular resolution of the PDOS projection to resolve the solvent PDOS as a function of distance between the solute center of mass and the ammonia nitrogen atoms (Figure~\ref{fig:dr-pdos}, main panels).
The uniformity of the resolved distribution along the distance axis is achieved by factoring out the probability density in this distance.
In an infinite system, this is proportional to $4\pi r^2 g(r)$, where $g(r)$ is the radial distribution function.
For our finite simulation cell, this quantity is shown in the bottom panel of Figure~\ref{fig:dr-pdos}; note the decay starting after $\sim$7~\AA\ that corresponds to half the length of the simulation box.
The distance resolution reveals a small systematic shift towards weaker electron binding energies in the proximity of the charged solute, up to 0.4~eV in the case of the $\mathrm{1e}$ peak. 
The origin of this effect can be attributed to the presence of the excess electron, since neutral benzene does not have a similar effect on liquid ammonia; its resolved peaks are essentially flat over the studied distance range (see Section~\ref{sec:additional-results} of the supplementary material).
The small magnitude of the perturbation of solvent one-electron levels by the solute can be used to justify the alternative method of spectrum resolution by subtraction of the neat solvent that is typically used in an experimental setting where a projection is not an option~\cite{Seidel2011/10.1021/jp203997p}.
The possible causes of the observed effect include the screening of the electrostatic interaction with the excess charge by the bulk solvent and are discussed in Section~\ref{sec:additional-results} of the supplementary material in terms of molecular clusters in open boundary conditions.
Additionally, we present a detailed validation of the required PDOS properties in Section~\ref{sec:additional-results} of the supplementary material. 

\section{Conclusions}
\label{sec:conclusions}

The reported analysis of the electronic structure of the solvated benzene radical anion in liquid ammonia complements the analysis of molecular geometry from our previous work and provides results that can be directly related to future experimental measurements of the system studied here.

The JT behavior of the solvated radical anion is analogous to that predicted for the idealized gas-phase species based on fundamental theory and symmetries.
The electronic state and its associated spin density correlates strongly with the dynamic distortion of the molecular geometry as it undergoes motion through the almost flat pseudorotation valley.
It thus turns out that the presence of the solvent is key to stabilize the studied system electronically but does not perturb it substantially from the perspective of the JT effect.
This sets the stage for possible experimental studies of the consequences of the JT effect on the molecular and electronic structure of the benzene radical anion which is not an option in the gas phase where the radical anion does not exhibit long-term stability.
However, such experiments would have to rely on ultrafast techniques so that the individual JT structures are observed rather than their high-symmetry average.

We quantified the solvent-induced stability of the benzene radical anion using accurate and computationally demanding condensed-phase G$_0$W$_0$ calculations performed on thermal geometries sampled from a hybrid DFT AIMD simulation.
We estimated the binding energy of the excess electron to be $-$2.34~eV relative to the vacuum level, clearly showing that the excess electron represents a bound quantum state in solution.
Moreover, the density of states obtained from such calculations predicts the complete valence electronic structure and thus provides a way to interpret future photoelectron spectroscopy measurements.

The present work showcases the descriptive power of accurate molecular simulations and detailed analysis of their outputs.
We captured subtle quantum effects in both the spatial and energy domains and obtained a detailed description of the solvated benzene radical anion in liquid ammonia, as well as a prediction of its electronic density of states that complements our previous prediction of the vibrational density of states.
The immediate next step lies in exploiting the synergy between the calculations reported here and future liquid photoelectron spectroscopy measurements. 
Referencing the results against the baseline of the solvated neutral benzene molecule further aids the interpretation of the anticipated experimental results.
This combination has the potential to experimentally corroborate the solvent-induced stability of the benzene radical anion.
One remaining issue is the computational description of the thermodynamic equilibrium between the benzene radical anion and solvated electrons that will provide additional insight into the experimentally observable chemical properties of the solvated benzene radical anion as well as an entryway to the theoretical exploration of the chemistry of the Birch reduction.

\section*{Supplementary Material}

Additional data analysis details, additional results concerning the spin density dimensionality reduction, the evaluation of the GMM clustering and the projected densities of states as well as a video file visualizing the evolution of spin density over the pseudorotating molecular structure of the benzene radical anion are presented as supplementary material.

\begin{acknowledgments}

K.B. acknowledges funding from the IMPRS for Many Particle Systems in Structured Environments.
This work was supported by the Project SVV 260586 of Charles University.
This work was partially supported by the OP RDE project (No. CZ.02.2.69/0.0/0.0/18\_070/0010462), International mobility of researchers at Charles University (MSCA- IF II).
P.J. is thankful for support from the European Regional Development Fund (Project ChemBioDrug no. CZ.02.1.01/0.0/0.0/16\_019/0000729) and acknowledges the Humboldt Research Award.
This work was supported by The Ministry of Education, Youth and Sports from the Large Infrastructures for Research, Experimental Development and Innovations Project ``IT4Innovations National Supercomputing Center -- LM2015070''.
The authors thank Hubert Beck and Tomáš Martinek for helpful comments on the manuscript.

\end{acknowledgments}

\section*{Data availability}

The data that support the findings of this study are available from the corresponding author upon reasonable request.

\section*{References}

%

\end{bibunit}


\clearpage

\setcounter{section}{0}
\setcounter{equation}{0}
\setcounter{figure}{0}
\setcounter{table}{0}
\setcounter{page}{1}

\renewcommand{\thesection}{S\arabic{section}}
\renewcommand{\theequation}{S\arabic{equation}}
\renewcommand{\thefigure}{S\arabic{figure}}
\renewcommand{\thetable}{S\arabic{table}}
\renewcommand{\thepage}{S\arabic{page}}
\renewcommand{\citenumfont}[1]{S#1}
\renewcommand{\bibnumfmt}[1]{$^{\rm{S#1}}$}

\title{Supplementary material for: \mytitle}
{\maketitle}

\begin{bibunit}

\section{Analysis details\label{sec:analysis-details}}

In the following paragraphs we discuss the details of the methods of data postprocessing and analyses that were used to obtain the key results presented in the main text.

\subsection{Spin density analysis details}

\begin{figure}[b!]
    \centering
    \includegraphics[width=\linewidth]{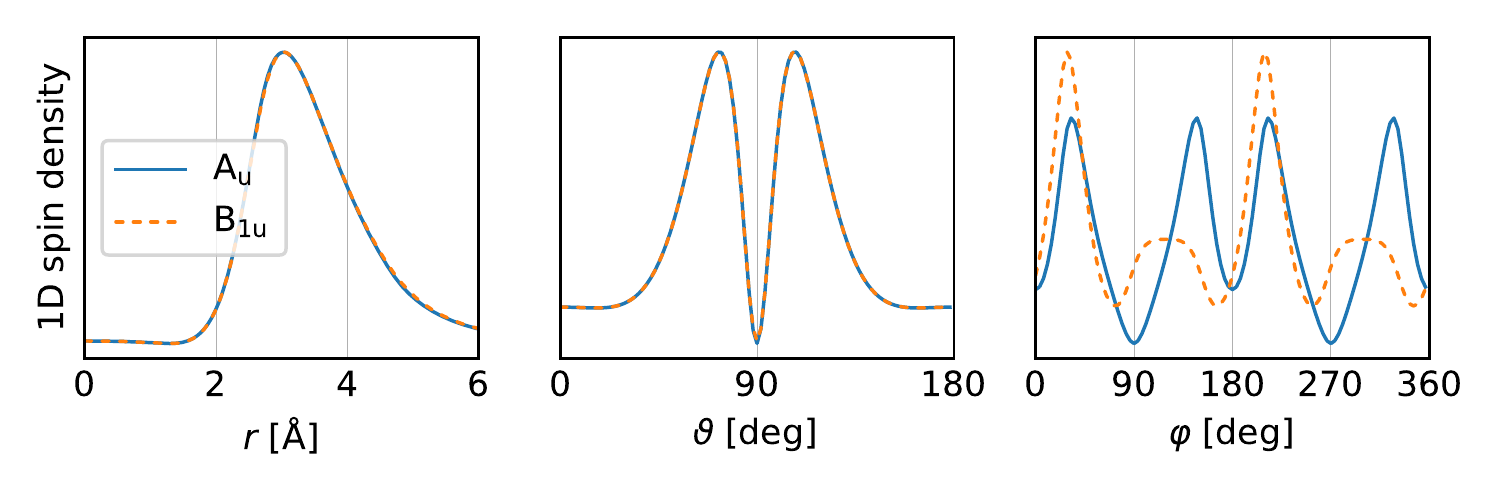}
    \caption{One-dimensional spin densities in the variables $r$, $\vartheta$, and $\varphi$.}
    \label{sifig:_1d}
\end{figure}

The 1D spin densities were obtained by partial integrations of the full, 3D spin density described using a local coordinate system as explained in the main text.
The radial density is calculated as
\begin{equation}
    \rho_\mathrm{s}(r) =
    \int_{0}^{\pi} \int_{0}^{2\pi} \dd \vartheta \dd \varphi \ \rho_\mathrm{s}(r, \vartheta, \varphi) \sin \vartheta, 
\end{equation}
the density in the azimuthal angle $\vartheta$ as
\begin{equation}
    \rho_\mathrm{s}(\vartheta) =
    \int_{0}^{r_{\mathrm{max}}} \int_{0}^{2\pi} \dd r \dd \varphi \ r^2 \rho_\mathrm{s}(r, \vartheta, \varphi)
\end{equation}
and, finally, the relevant density in the polar angle $\varphi$ as
\begin{equation}
    \rho_\mathrm{s}(\varphi) =
    \int_{0}^{\pi} \int_{0}^{r_{\mathrm{max}}} \dd \vartheta \dd r \ \rho_\mathrm{s}(r, \vartheta, \varphi) r^2 \sin \vartheta. 
\end{equation}
The value of $r_{\mathrm{max}}$ is fixed by the maximal available extent of the volumetric data.
All of these 1D spin densities for the ideal structures are shown in Figure~\ref{sifig:_1d}.
Clearly, only $\rho_\mathrm{s}(\varphi)$ (right panel) distinguishes between the $\mathrm{A_{u}}$ and $\mathrm{B_{1u}}$ states; the remaining two 1D spin densities show perfect overlap for the two states in question.

\begin{figure}[tb!]
    \centering
    \includegraphics[width=\linewidth]{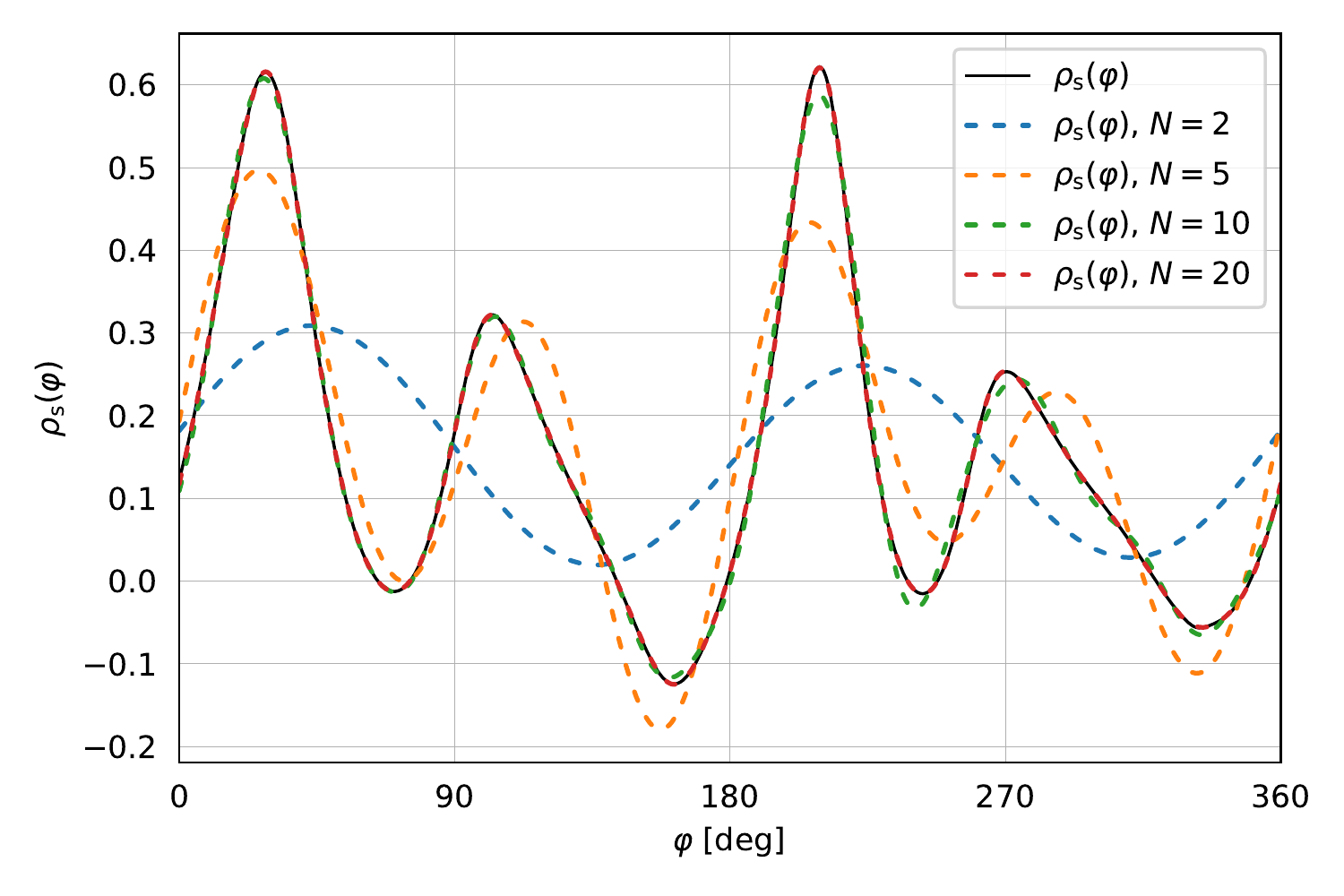}
    \caption{Reconstructed sample of a thermal 1D spin density in $\varphi$ by decomposition into the Fourier basis of size $N =$ 2, 5, 10 and 20.}
    \label{sifig:fourier}
\end{figure}

The $P = 2\pi$-periodic densities $ \rho_\mathrm{s}(\varphi)$ were further decomposed into the Fourier basis of harmonic functions $\exp(i n 2 \pi \varphi / P), n \in \mathbb{Z}$, collecting the complex coefficients of all degrees $n$
\begin{equation}
    c_n = \frac{1}{P} \int_0^P \dd\varphi \ \rho_\mathrm{s}(\varphi) \exp\left( - \frac{i n 2 \pi \varphi}{P} \right)
\end{equation}
into a $(2N+1)$-dimensional vector $\mathbf{c}$. 
Already $N=20$ was observed to very closely represent the original 1D densities: an example featuring a random sample from the thermal ensemble of 1D spin densities and its Fourier decomposition into harmonic bases of varying size $N$ is shown in Figure~\ref{sifig:fourier}. 

\section{Additional results\label{sec:additional-results}}

Here, we present additional results that complement and validate the main text results.
These include validations of the GMM clustering as well as of the PDOS projections and complementary analyses of the PDOS of both the solute and the solvent that provide data to support the discussion in the main text.

\subsection{Validation of GMM clustering}

The values in Table~\ref{tab:clustering} reveal the fact the distances between closest cluster centers are almost an order of magnitude higher than the standard deviations of the individual clusters and thus to a clean clustering of the data in question.

\begin{table}[htb!]
    \centering
    \begin{tabular}{c | c c c c c c}
    \toprule
    Cluster & 1 & 2 & 3 & 4 & 5 & 6 \\
    \midrule
    $d$ [rad$^{-1}$] & 0.350 & 0.334
    & 0.345 & 0.342 & 0.350 & 0.334  \\
    $\sigma$ [rad$^{-1}$] & 0.0554 & 0.0554 & 0.0555 & 0.0551 & 0.0552 & 0.0552
    \end{tabular}
    \caption{Clustering characteristics. 
    First row: the Euclidean distance ($d$) of each cluster to its nearest neighbor.
    Second row: Standard deviation ($\sigma$) of each cluster.}
    \label{tab:clustering}
\end{table}

\begin{figure}[tb!]
    \centering
    \includegraphics[width=\linewidth]{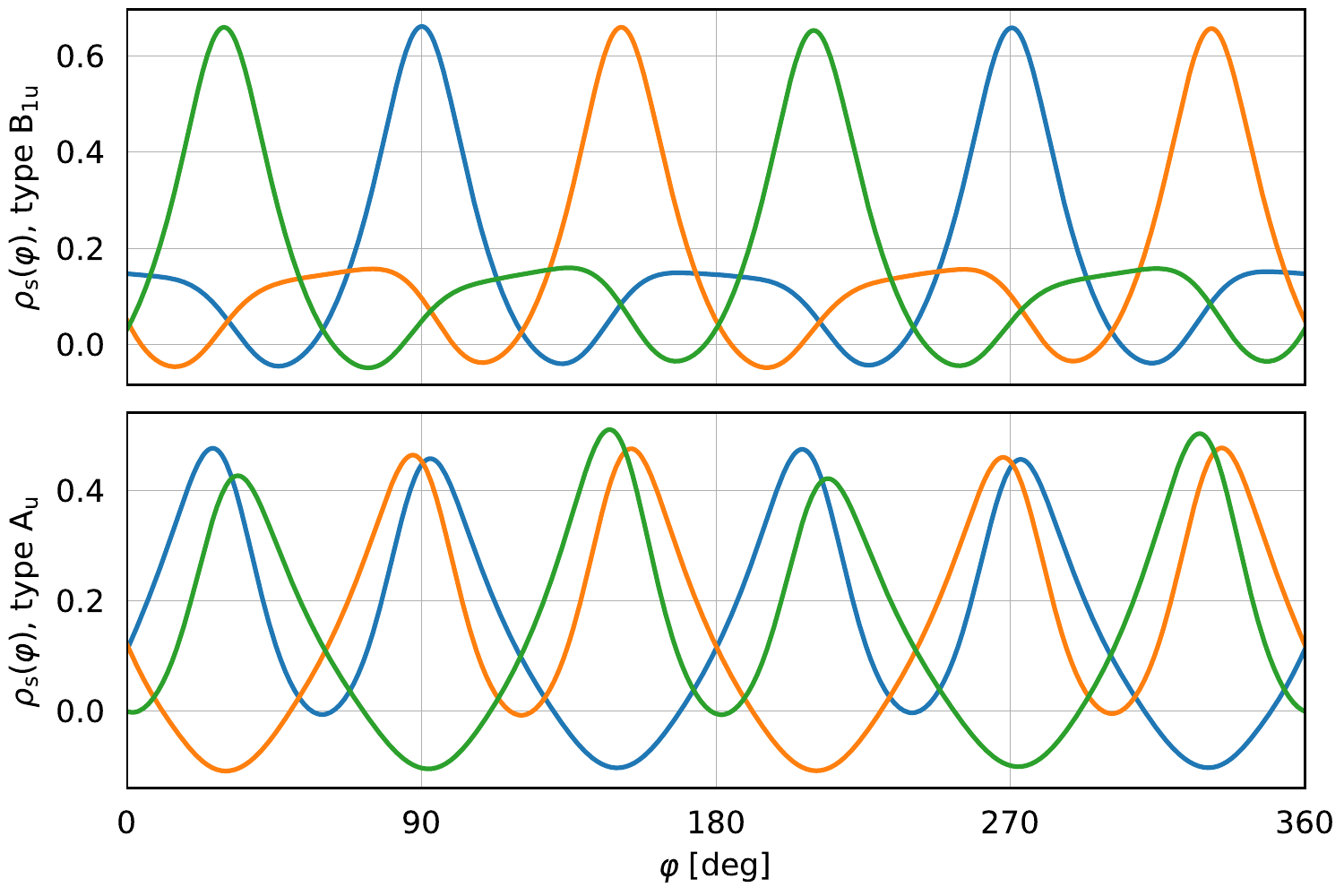}
    \caption{One-dimensional spin densities obtained by Fourier synthesis of the means of the GMM clusters.
    Different colors correspond to different clusters.
    }
    \label{sifig:GM_means}
\end{figure}

Next, we verify that the obtained clusters indeed represent the spin density shapes expected on physical grounds.
This is achieved by reconstructing new (\textit{i.e.}, previously not existing in the AIMD data) 1D spin densities $\Bar{\rho_{\mathrm{s}}}_i(\varphi)$ from the cluster means $\Bar{\mathbf{c}}_i$ using the Fourier synthesis
\begin{equation}
    \Bar{\rho_{\mathrm{s}}}_i(\varphi) =
    \sum_{n=-N}^{N} \Bar{c}_{i,n} \exp\left(\frac{i n 2 \pi \varphi}{P} \right).
\end{equation}
These mean spin densities are shown in Figure~\ref{sifig:GM_means}.
Clearly, three curves of the $\mathrm{A_u}$ type and three curves of the $\mathrm{B_{1u}}$ type are obtained which respect the 120$^\circ$ shifts in $\varphi$ due to the pseudorotation.
These final means are fully invariant to the repetition of the clustering staring from different random initial mean estimates.

\subsection{Nuclear GMM clustering performance}

\begin{figure}[b!]
    \centering
    \includegraphics[width=\linewidth]{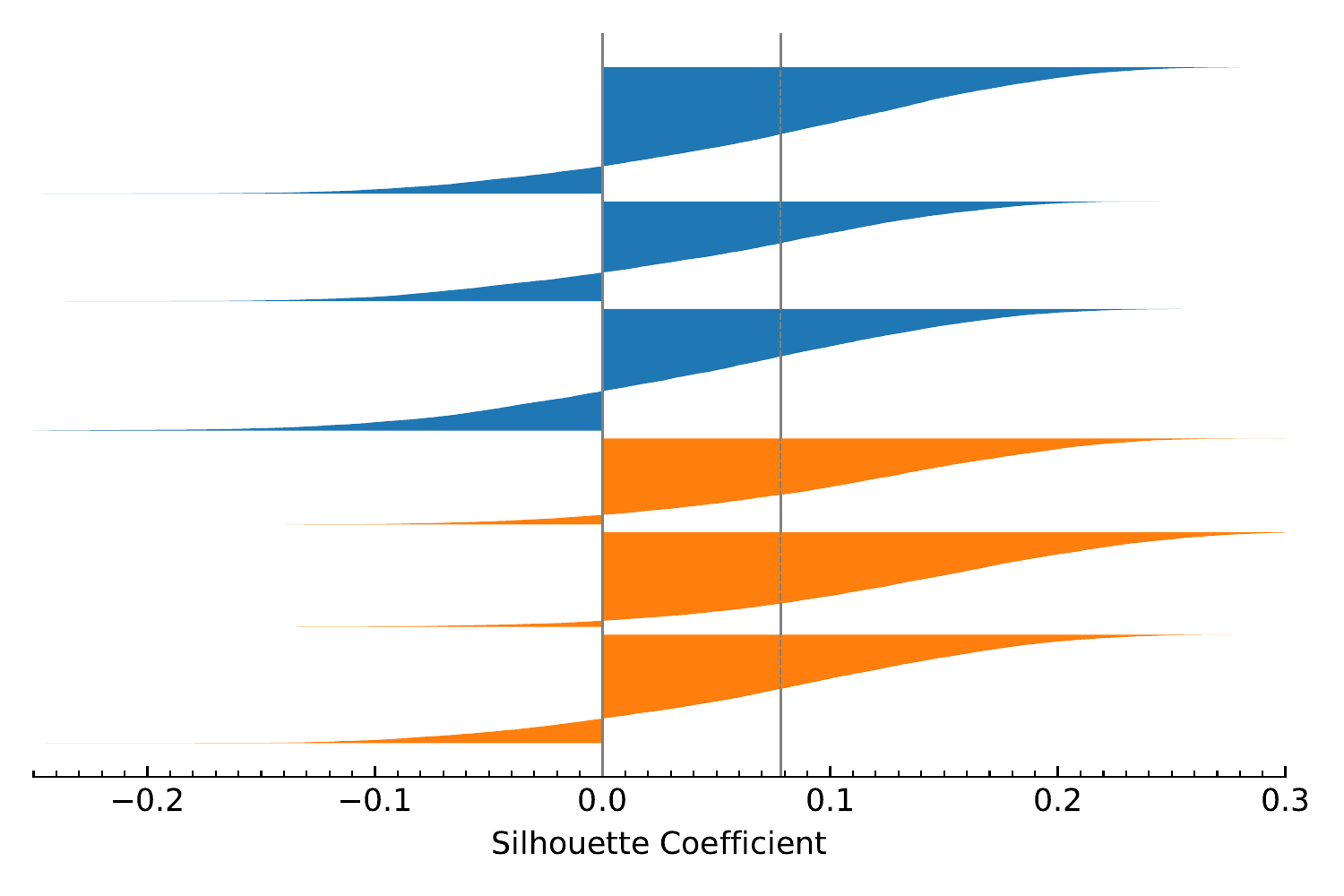}
    \caption{Silhouette coefficients for the clustering of nuclear geometries.
    }
    \label{sifig:nuclear-silhouettes}
\end{figure}

The silhouette plot showing the quality of clustering of nuclear geometries is shown in Figure~\ref{sifig:nuclear-silhouettes}.

\subsection{Validation of PDOS properties}

\begin{figure}[tb!]
    \centering
    \includegraphics[width=\linewidth]{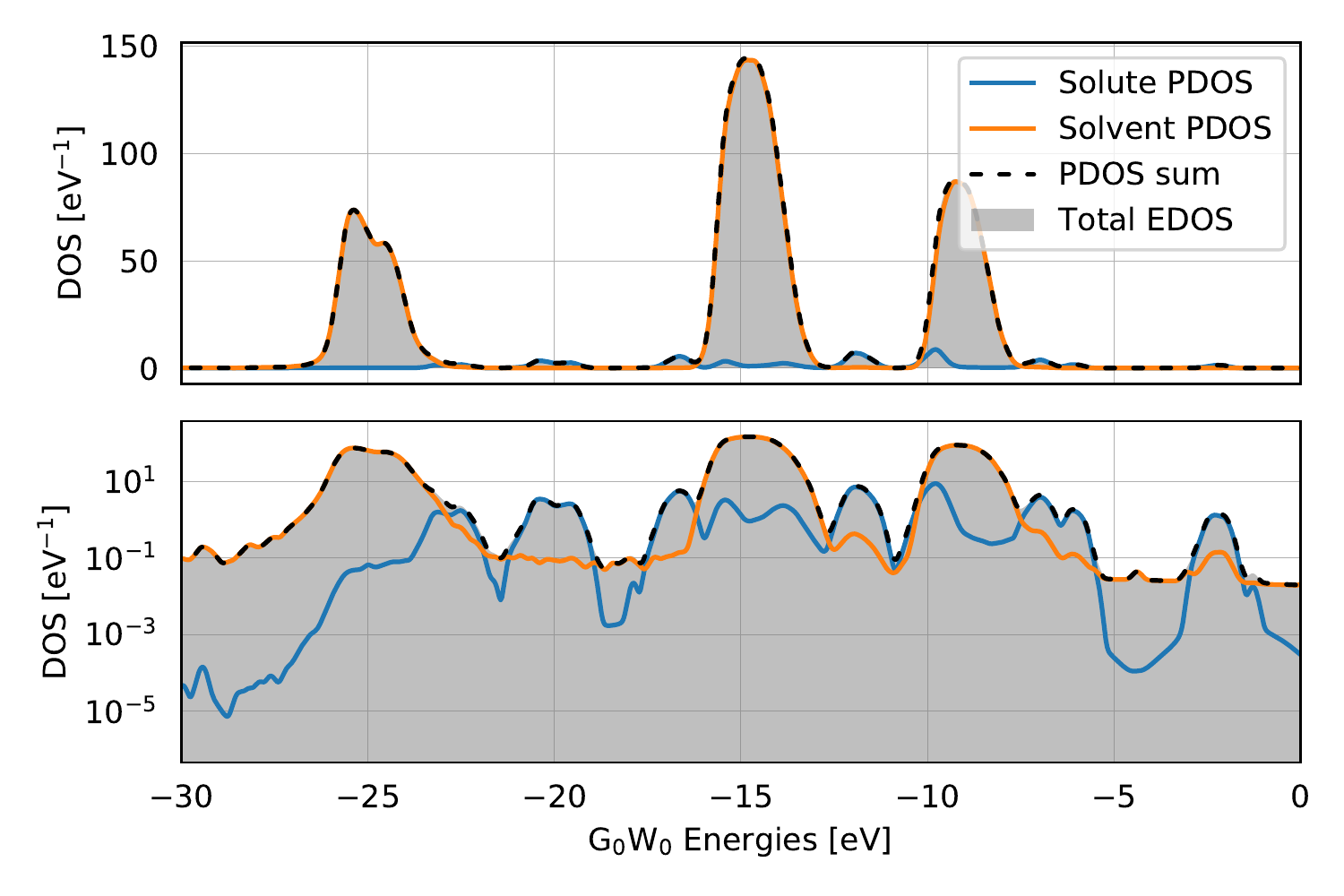}
    \caption{Summation of solute and solvent PDOS for the benzene radical anion.
    The top panel shows that the sum (black, dashed) of solute (blue) and solvent (orange) PDOS  reconstructs the total EDOS (gray shading) perfectly.
    The bottom panel shows the same situation in logarithmic scale to bring out the low-intensity details.}
    \label{sifig:sum-PDOS}
\end{figure}

The PDOS theory presented in Section II of the main text guarantees though Equation~3 that a set of PDOS curves for mutually exclusive molecular subsystems that span the whole system must sum exactly to the total EDOS.
This property is clearly illustrated for the system containing the benzene radical anion in Figure~\ref{sifig:sum-PDOS} where the sum of the solute and solvent PDOS perfectly recreate the total EDOS.
An identical situation can be demonstrated for the neutral system which is, however, not shown explicitly.

\begin{figure}[b!]
    \centering
    \includegraphics[width=\linewidth]{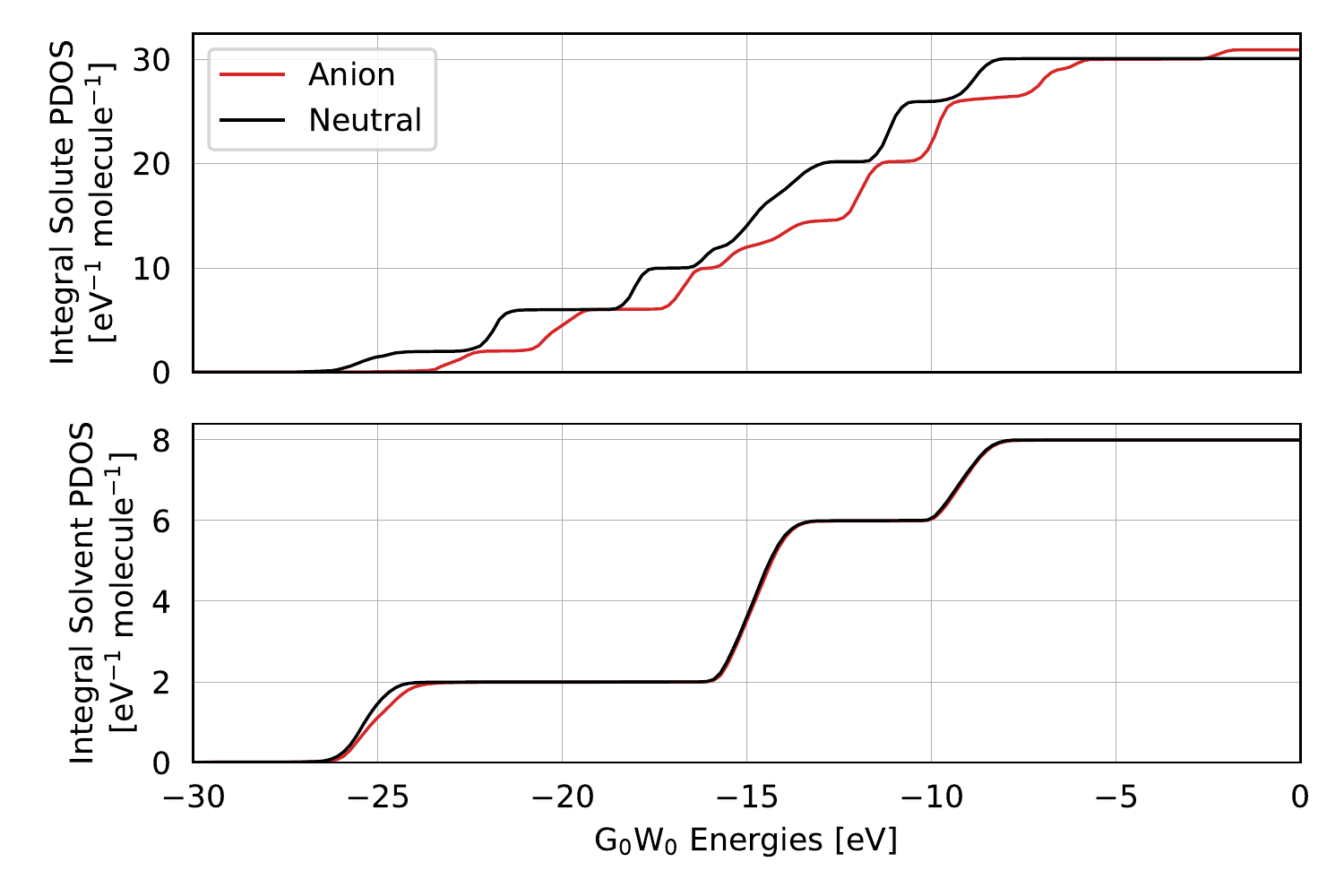}
    \caption{Cumulative integral of the solute (top) and solvent (bottom) PDOS for the anionic (red) and neutral (black) systems.}
    \label{sifig:cumsum-PDOS}
\end{figure}

However, the sole fact that the PDOS curves sum up to the total EDOS does not say anything about the quality of the individual projections.
This strongly depends on the coupling between the subsystems and the character of the KS orbitals that enter the calculation of the weights, in particular on their spatial localization.
The inspection of the clarity of the projection, \textit{i.e.}, if it is not contaminated by diffuse states extending from another molecular subsystem, relies on the cumulative integration of the PDOS curves.
If the projection is indeed meaningful in terms of molecularity, then the individual PDOS curves should integrate to the number of electrons that would be expected for an isolated species.
For the neutral and anionic solutes, this is 30 and 31 explicit electrons since the core $1s$ levels are represented by pseudopotentials in our calculations.
For the solvent, one can expect eight explicit electrons per molecule.
The cumulative integrals of the PDOS curves shown in Figure~\ref{sifig:cumsum-PDOS} show excellent agreement with these predictions.

\subsection{Solute contribution obtained by subtraction}

\begin{figure}[tb!]
    \centering
    \includegraphics[width=\linewidth]{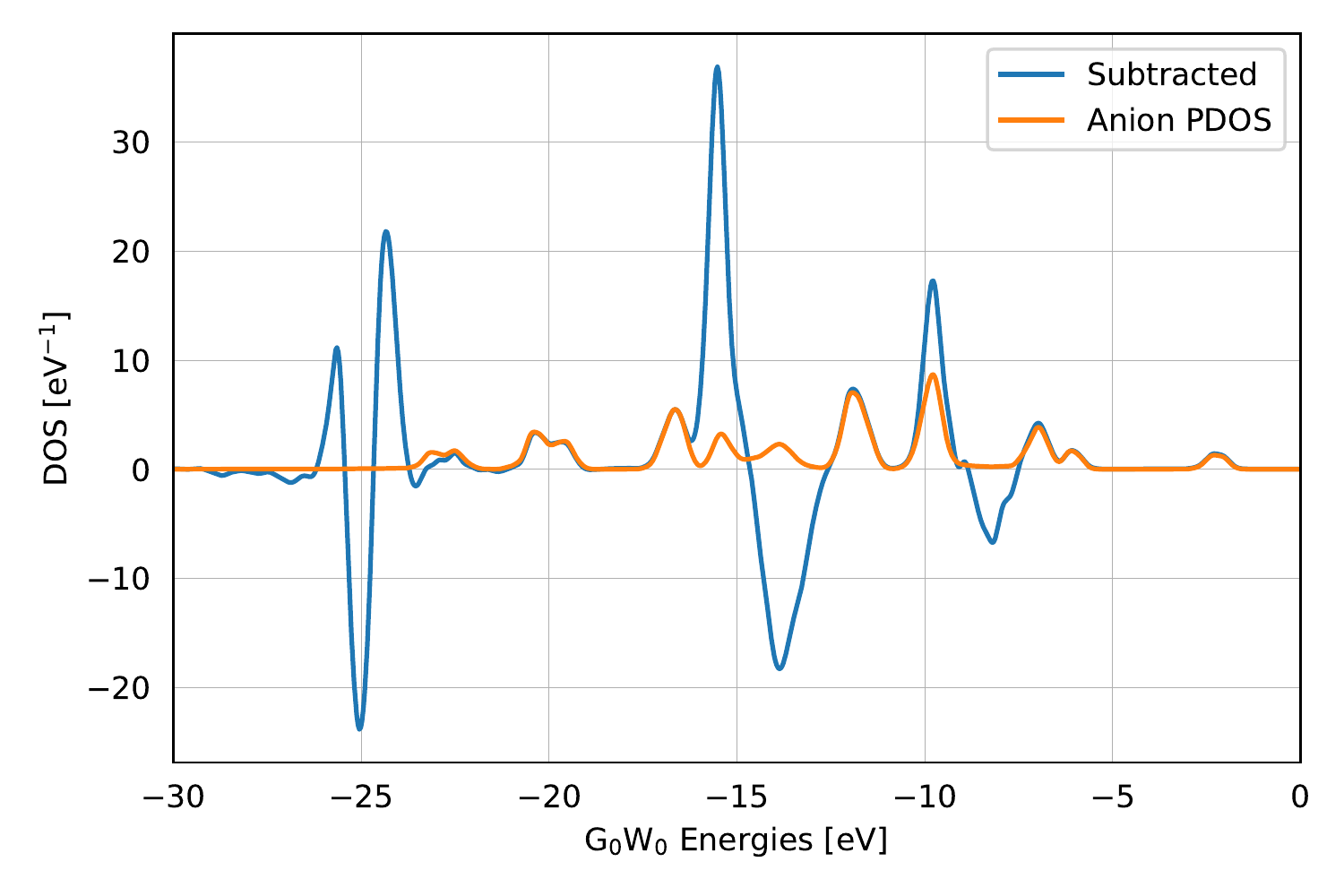}
    \caption{Comparison of projection and subtraction for the solute contribution.
    The contribution of the benzene radical anion to the total EDOS of the system obtained by subtraction of the neat liquid ammonia EDOS from the total EDOS of the system including the solute is shown in blue.
    The solute PDOS is shown in orange for reference.}
    \label{sifig:subtraction}
\end{figure}

An alternative way to isolate the solute contribution from the total EDOS that is more oriented towards the experimental approach relies on subtraction of the neat liquid ammonia EDOS from the total EDOS of the system including the solute.
The result of this subtraction is shown and compared to the solute PDOS in Figure~\ref{sifig:subtraction}.
The subtraction produces an acceptable outcome where the solute states do not overlap with the ones of the solvent.
However, in those areas where they do, the noise arising from the seemingly subtle differences between the solvent in the total EDOS and the neat one leads to overpowering artifacts in the result of the subtraction.
Clearly, this effect is concentration dependent and the more concentrated the solute would be and thus its signal stronger, the less of an issue it presents.
The differences between the EDOS of the system with the solvated benzene radical anion and the neat solvent have several possible origins.
The oscillatory character of the noise in the subtracted signal suggests a horizontal shift of the peaks in question, which likely combines the physical shift caused by the different chemical environments in the compared systems, the numerical details of the individual G$_0$W$_0$ calculations as well as the numerical details pertaining to the alignment of the absolute energies.
In particular, the latter has a profound effect on the subtraction artifacts, but also has a significant amount of numerical freedom including the calculation of single-peak means and the choice of the reference density of states, all without a clearly correct option.
Furthermore, the previous simulations of the neat solvent were performed at the experimental density, while the simulations with the solute had the estimated molar volume of the solute added.
It is impossible to ensure that the density of the solvent in the two cases is exactly identical, which can further contribute to minor differences between the EDOS of the solvent in the two cases that we are trying to subtract.
In this light, the PDOS offers a superior computational treatment since it does not suffer from these issues. 
However, it must be expected that the experimental measurement will have to deal with subtraction artifacts in some form and to some extent.

\subsection{Geometric insight into the solvent distance-resolved PDOS}

\begin{figure}[b!]
    \centering
    \includegraphics[width=\linewidth]{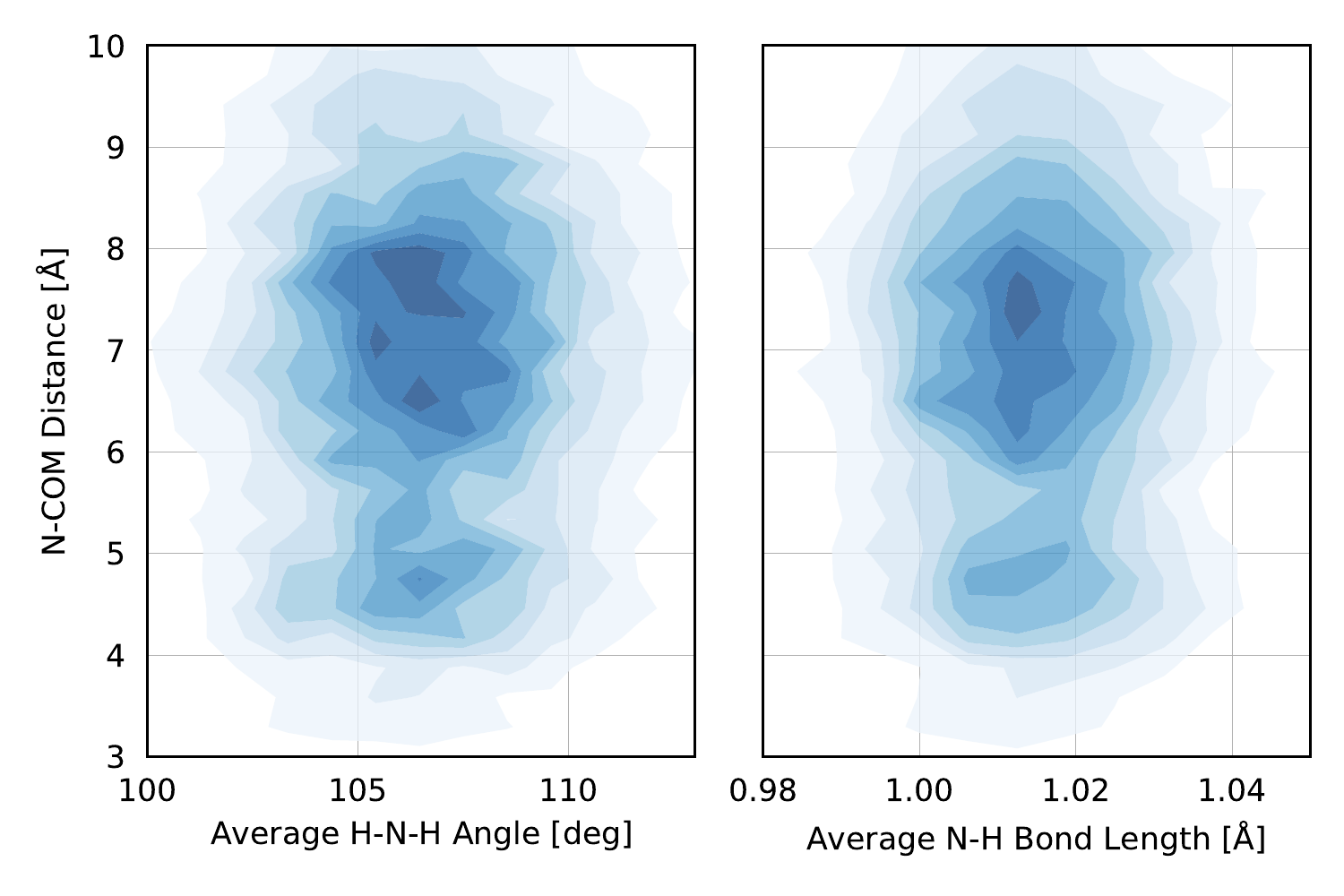}
    \caption{Correlation between the distance of a solvent nitrogen form the benzene radical anion center of mass and the average H--N--H angle (left) and average N--H bond length (right).
    The blue color scale shows the normalized probability distribution over the pairs of coordinates in question.}
    \label{sifig:geometry}
\end{figure}

In this section, we demonstrate that the effect observed in the one-electron levels of the solvent in proximity of the solute demonstrated in the main text in Figure~\ref{fig:dr-pdos} is a first-order effect caused directly by the solute perturbing the solvent electron rather than a geometry-mediated second-order effect.
In particular, we observe no correlation between the distance of the solvent from the solute center of mass and the average molecular H--N--H angle and N--H bond length and thus no molecular geometry change is expected for the molecules in proximity of the solute.
These correlations are shown as bivariate probability densities in Figure~\ref{sifig:geometry} and correspond to the average bending and stretching of the ammonia solvent molecules.
It thus seems that the perturbation of the solvent one-electron levels in molecules that are in close contact with the charged solute is caused directly by the solute affecting the solvent electrons.

\subsection{Solvent One-electron Levels Outside of the Bulk Solution}

\begin{figure}[tb!]
    \centering
    \includegraphics[width=\linewidth]{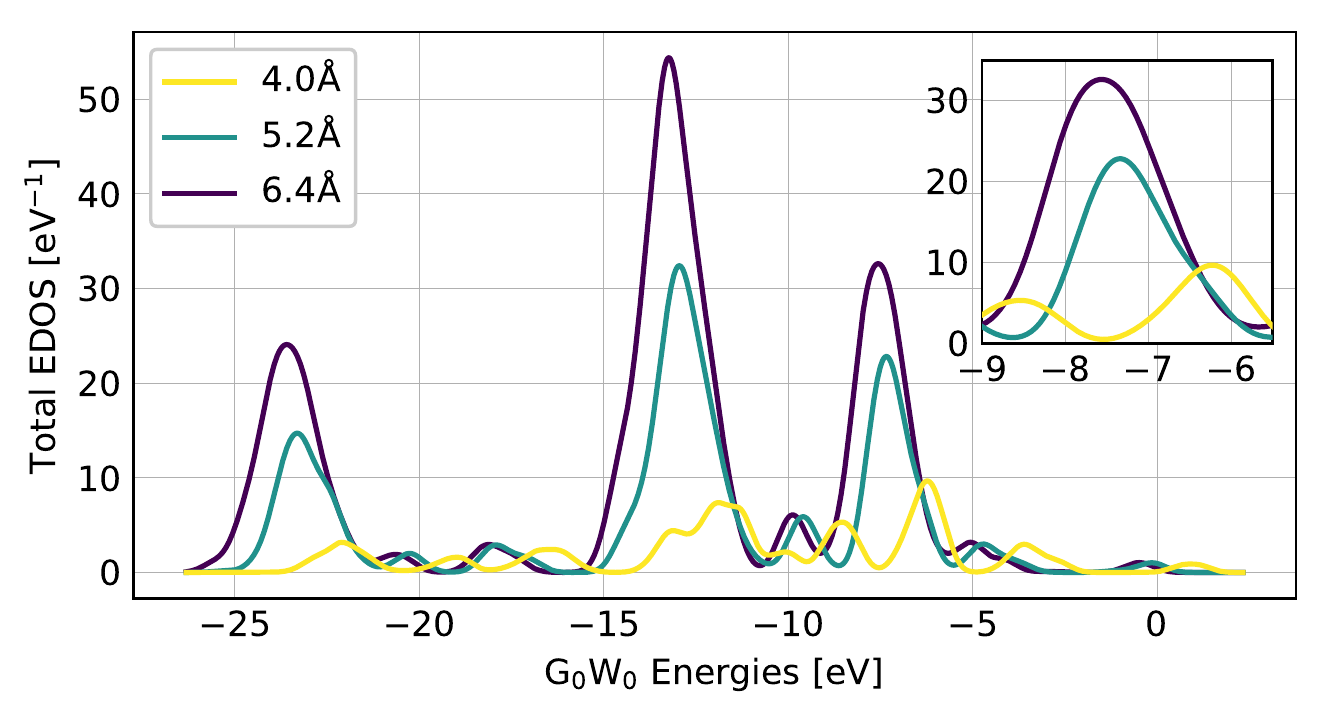}
    \caption{Open boundary conditions G$_0$W$_0$ EDOS of molecular clusters carved out of the bulk AIMD structures.
    The inset shows the first solvent peak in detail.}
    \label{sifig:clusters}
\end{figure}

It might be expected that a major component to this effect is the electrostatic effect of the excess charge.
In the bulk solution, the electrostatic contribution must be noticeably screened by the dielectric environment provided by liquid ammonia.
Thus, a further level of qualitative insight into the perturbation of the solvent one-electron levels is achieved by carving out small molecular clusters including just the closest solvent molecules and recalculating the G$_0$W$_0$ in open boundary conditions.
Note that the open boundaries directly provide absolute binding energy values and no additional alignment is necessary.
This is shown in Figure~\ref{sifig:clusters} for spherical clusters of various radii.
Here, the solvent peaks are shifted more significantly by +1.5 to +3.0~eV towards lower binding energies in comparison to the +0.4~eV maximal bulk shift of the $\mathrm{1e}$ peak.
A detailed insight into the problem of benzene-radical-anion--ammonia clusters is provided by hybrid DFT calculations in the gas phase~\cite{Kostal2021/10.1021/acs.jpca.1c04594}.

\subsection{Neutral benzene solvent PDOS}

\begin{figure}[tb!]
    \centering
    \includegraphics[width=\linewidth]{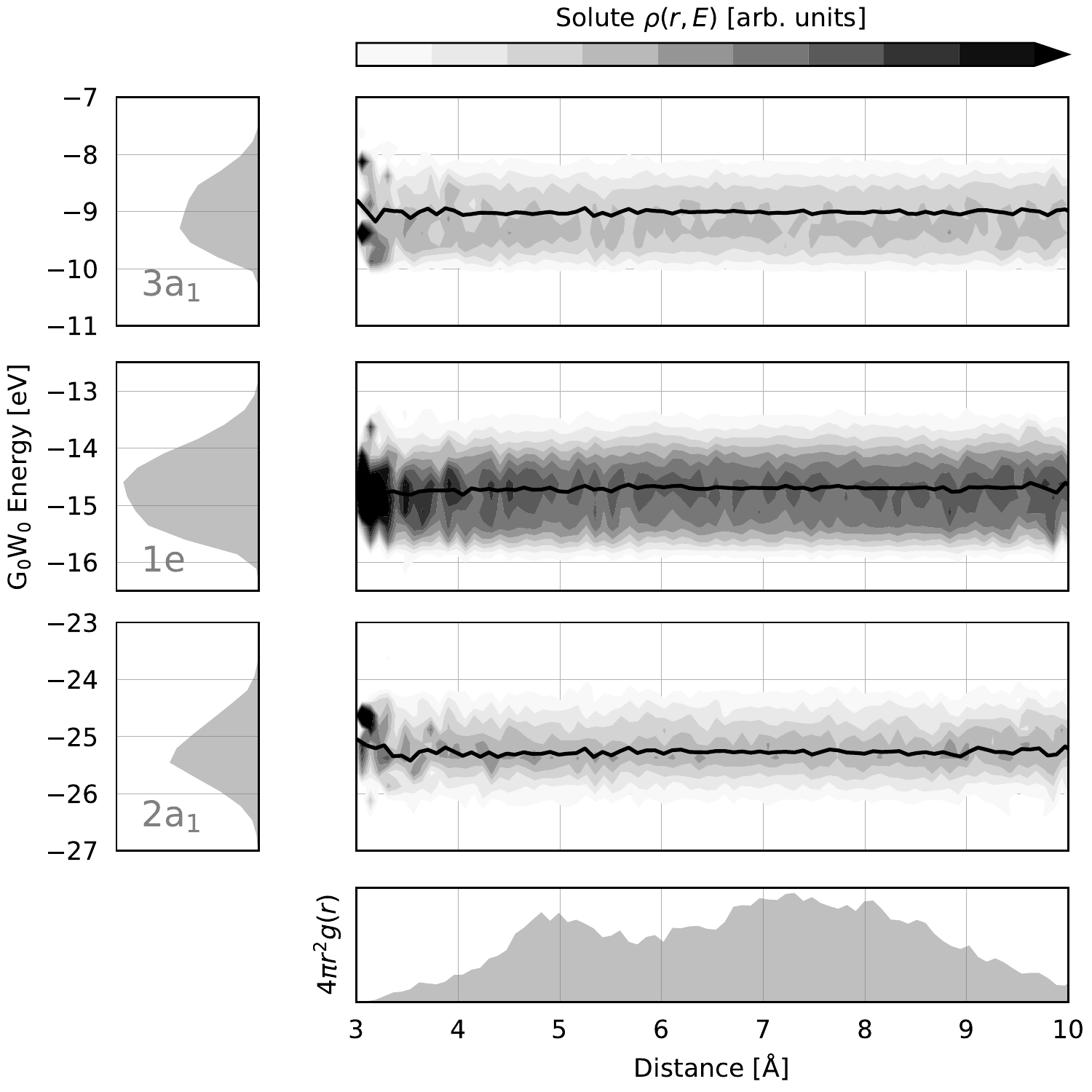}
    \caption{Electronic density of states projected on the solute subspace and resolved as a function of distance from the center of mass of the neutral benzene.
    Additional details of the plots are identical with the main text Figure~\ref{fig:dr-pdos}.}
    \label{sifig:dr-pdos-neutral}
\end{figure}

Figure~\ref{sifig:dr-pdos-neutral} shows the neutral counterpart of the anion Figure~\ref{fig:dr-pdos} of the main text.
Note that in this case the distance-resolved peaks show no visible effect in solvent localized in proximity to the solute.

\section{Description of video files\label{sec:SI_videos}}

To visualize the pseudorotation of the electronic structure on top of a pseudorotating molecular geometry of the benzene radical anion, we present the video file \texttt{excess\_electron\_pseudorotation.mp4}.
This captures the pseudorotation of the molecular geometry obtained as an artificial motion of the atomic positions along the two relevant orthonormal vibrational modes $Q_x$ and $Q_y$ defined in the main text.
The ideal circular pseudorotation path is achieved by the following propagation
\begin{equation}
\begin{split}
    & Q_x(t) = A \cos \omega t \\
    & Q_y(t) = A \sin \omega t,
\end{split}
\end{equation}
where $t$ is time parameter, $\omega$ is the vibrational frequency of the degenerate mode pair ($\hbar\omega$ = 1747~cm$^{-1}$) and the amplitude of motion $A =$ 0.25 \textit{a.u.} which follows from the natural position of the pseudorotational minimum path in the simulated data (Figure~\ref{fig:JT}, main text).
On top of the molecular trajectory we calculated the electronic structure evolution at the revPBE0-D3/TZVP level of theory.
The spin density contours are shown at the values of $\pm$0.006~\AA$^{-3}$ and $\pm$0.0025~\AA$^{-3}$, colored in green and purple for the positive and negative parts respectively.
The image files were rendered using the Tachyon ray tracer~\cite{Stone1998} embedded in the VMD molecular visualization software~\cite{Humprhey1996/10.1016/0263-7855(96)00018-5}.
For smooth animation, the propagation time step of 0.1255~fs was used for the discrete parametrization of the path, which results in the video playback speed of 5.07~s/pseudorotation period with the used 30~fps frame rate.

\end{bibunit}

\end{document}